\documentclass[iop,apj]{emulateapj} 

\slugcomment{Draft v5.4}

\shorttitle{Extreme Debris Disks in Time Domain}
\shortauthors{Meng et al.}

\begin{document}
\title{Planetary Collisions outside the Solar System: Time Domain Characterization of Extreme Debris Disks}

\author{Huan Y. A. Meng\altaffilmark{1,2}, Kate Y. L. Su\altaffilmark{2}, George H. Rieke\altaffilmark{1,2}, Wiphu Rujopakarn\altaffilmark{1,3,4}, Gordon Myers\altaffilmark{5}, Michael Cook\altaffilmark{5}, Emery Erdelyi\altaffilmark{5}, Chris Maloney\altaffilmark{5}, James McMath\altaffilmark{5}, Gerald Persha\altaffilmark{5}, Saran Poshyachinda\altaffilmark{6}, Daniel E. Reichart\altaffilmark{7}}
\altaffiltext{1}{Lunar and Planetary Laboratory, Department of Planetary Sciences, University of Arizona, 1629 East University Boulevard, Tucson, AZ 85721}
\altaffiltext{2}{Steward Observatory, Department of Astronomy, University of Arizona, 933 North Cherry Avenue, Tucson, AZ 85721}
\altaffiltext{3}{Department of Physics, Faculty of Science, Chulalongkorn University, 254 Phayathai Rd, Pathumwan, Bangkok 10330, Thailand}
\altaffiltext{4}{Kavli Institute for the Physics and Mathematics of the Universe (WPI), Todai Institute for Advanced Study, University of Tokyo, 5-1-5 Kashiwanoha, Kashiwa, 277-8583, Japan}
\altaffiltext{5}{American Association of Variable Star Observers, 49 Bay State Road, MA 02138}
\altaffiltext{6}{National Astronomical Research Institute of Thailand (Public Organization), Ministry of Science and Technology, 191 Siriphanich Bldg, Huay Kaew Rd, Muang District, Chiang Mai 50200, Thailand}
\altaffiltext{7}{Department of Physics and Astronomy, Campus Box 3255, University of North Carolina at Chapel Hill, Chapel Hill, NC 27599}
\email{hyameng@lpl.arizona.edu}

\begin{abstract}
Luminous debris disks of warm dust in the terrestrial planet zones around solar-like stars are recently found to vary, indicative of ongoing large-scale collisions of rocky objects. We use {\it Spitzer} 3.6 and 4.5 $\micron$ time-series observations in 2012 and 2013 (extended to 2014 in one case) to monitor 5 more debris disks with unusually high fractional luminosities (``extreme debris disk''), including \objectname{P1121} in the open cluster \objectname{M47} (80 Myr), \objectname{HD 15407A} in the \objectname{AB Dor moving group} (80 Myr), \objectname{HD 23514} in the \objectname{Pleiades} (120 Myr), \objectname{HD 145263} in the \objectname{Upper Sco} Association (10 Myr), and the field star \objectname{BD+20 307} ($\gtrsim$1 Gyr). Together with the published results for \objectname{ID8} in \objectname{NGC 2547} (35 Myr), this makes the first systematic time-domain investigation of planetary impacts outside the solar system. Significant variations with timescales shorter than a year are detected in five out of the six extreme debris disks we have monitored. However, different systems show diverse sets of characteristics in the time domain, including long-term decay or growth, disk temperature variations, and possible periodicity.
\end{abstract}



\section{Introduction}

With temperatures typically between $\sim$30 and 300 K, debris disks are gas-poor analogs of the asteroid belt and Kuiper belt in the solar system \citep{bac93,wya08}, emerging after gas-dominant primordial protoplanetary disks dissipate and representing the relics of planetary formation and evolution. Debris disks have been found around stars at nearly all stages of stellar evolution, from pre-main sequence to mature main sequence \citep{wya08} and even white dwarf \citep{far09}. They are dynamically maintained by collisional cascades of small bodies and dust down to the blowout sizes of their host stars \citep{wya08}. The fractional luminosities, defined as the ratios of disk to stellar luminosity, of most debris disks are $<10^{-4}$. The typical values are consistent with simple models in which the debris mass is initially proportional to that of its ancestral protoplanetary disk and decays through pseudo-equilibrium collisional cascades \citep{wya07}. Over the stellar lifetime, debris disks undergo a decay in infrared excess with a timescale of hundreds of Myr \citep{rie05,su06} due to the erosion of the large bodies participating in the dust production \citep{gas13}. Most dust particles in classical debris disks have blackbody-like radiation, leading to featureless spectral energy distributions (SED) for most debris disks.

Recent discoveries revealed a new class of debris disk, ``extreme debris disk'', characterized by very high fractional luminosities ($\gtrsim$10$^{-2}$) that are orders of magnitude above the upper limit of equilibrium evolution. Their evolution cannot be dominated by collisional cascades \citep{men12,mel12,sch13}. Instead, extreme debris disks probably reflect recent large planetary impacts and their immediate aftermath \citep{men14}. Many of them show spectral features of very fine dust particles, including some cases with silica dust or even vapor \citep[e.g.,][]{bal09,lis09} that are from hypervelocity collisions or shocks \citep{mor14}. Some extreme debris disks are known to have variable emission over timescales of order 1 year and shorter \citep{men12,mel12}. In addition, with only one known exception (\objectname{BD+20 307}, $\gtrsim$1 Gyr), all extreme debris disks are found around young stars in the age range of $\sim$10 to 200 Myr. This coincides with the era of the final stage of terrestrial planet formation in the solar system, which is characterized by massive collisions \citep[e.g.,][]{ken06,mor10,rig11,ste12,cha13,ray14} up to the scale of the Moon-forming impact \citep{can04,cuk12,can12}.

One of the prototypes of extreme debris disks, \objectname{ID8} (2MASS J08090250-4858172) in the open cluster \objectname{NGC 2547}, was found to be variable on roughly a yearly timescale based on 3 epochs of 24 $\micron$ observations \citep{men12} made by the Multiband Imaging Photometer for {\it Spitzer Space Telescope} \citep[MIPS,][]{rie04}. In early 2013, we monitored its near-infrared variations intensively and detected an outburst of disk emission \citep{men14}. In this work, we analyze the near-infrared monitoring of five additional extreme debris disks from 2012 to 2014 by IRAC \citep{faz04} at 3.6 and 4.5 $\micron$. All but one of the stars have coordinated optical monitoring from the ground. None shows significant stellar activity, as detailed in the appendix. Four of these debris disks have varied during our monitoring, demonstrating that the time domain is a useful dimension for the study of terrestrial planet formation. We will introduce the observations and data reduction in \S2, analyze and discuss each light curve along with the mid-infrared SEDs in \S3, and explore the implications of these results in \S4.

\section{Observations and Data Analysis}

\subsection{Targets}

Our targets were selected based on several criteria to maximize the likelihood of seeing debris disk variability with {\it Spitzer}. First, we focus on debris disks with mid-infrared properties indicating fractional luminosities, $f_d = L_{disk}/L_*$ greater than several times $10^{-3}$. Second, the samples are restricted to dwarf stars with spectral types between F and mid-K to focus on terrestrial planet formation around solar-like stars. Third, the targets of interest must have mid-infrared spectra that show prominent features of fine silica or silicate dust at 9 to 10 $\micron$. Fine dust (typically sub-$\micron$-sized) should have short dwell time around the stars and is likely indicative of recent collisions. Finally, the disks must have significant excesses at 3.6 and 4.5 $\micron$ (roughly $T_{disk} \gtrsim 400$ K) that yield a sufficient signal-to-noise ratio (S/N) with IRAC on the warm {\it Spitzer} mission.

As expected, these conditions naturally selected debris disks around stars younger than 200 Myr, with only one exception (BD+20 307). Our final sample set includes 6 stars\footnote{At least two of the stars, HD 23514 and BD+20 307, have companions within our photometric aperture.}, listed in Table~\ref{spitzer} except for ID8 that has been analyzed in previous works \citep{men14}. 

\subsection{{\it Spitzer} Observations and Data Reduction}

New {\it Spitzer} observations analyzed in this work were conducted under programs 80260 and 90192. In a typical pattern due to {\it Spitzer}'s pointing restrictions, all five selected sources have two visibility windows each year, and each visibility window lasts 30-50 days. For each target, a fixed sampling frequency of 1 AOR every 5 days was used throughout all observations. The only exception is BD+20 307, a spectroscopic binary system with two G0V stars, for which we used two cadences, including the regular cadence of 1 AOR every 4 days and two periods of high cadence of 1 AOR per day to probe the gravitational effects of the stellar orbital period on the disk. As a general strategy, we used several cycling dithering positions for each AOR to average the intrapixel sensitivity variations of the IRAC detector array. The time coverage and sampling cadence for each target are listed in Table~\ref{spitzer}.

The {\it Spitzer} data were first processed with IRAC pipeline S19.1.0 by the {\it Spitzer} Science Center (SSC). As found in previous works \citep{men14}, post-BCD mosaics are prone to WCS misalignment that can lead to erratic photometry. In this work, all our photometry was based on the BCD images. The BCD images come with a scale of 1.22\arcsec\ pixel$^{-1}$. An aperture of 3 pixels in radius and a sky annulus between radii of 12 and 20 pixels were used with aperture correction factors of 1.112 and 1.113 for 3.6 and 4.5 $\micron$, respectively. Mapping distortions of the images were corrected with files provided by the SSC. HD 15407A is so bright that the system output at 3.6 and 4.5 $\micron$ would be saturated in the full-array mode; it was observed in the sub-array mode. For the sub-array data, we performed photometric extraction on the median combined BCD data cube ({\it sub2d.fits} files) provided by the SSC. The same aperture and sky annulus settings were used for the sub-array mode data, and the final photometry of each observation was obtained by weighted averaging of the results for all dither positions. The weights are chosen to be the squared inverse of the nominal photometric error for each of the dithered frames. Since we have many dither positions (or repeats in the sub-array mode), in computing the average we also exclude individual BCD exposures in which the sources are too close to the edge of the detector array, as well as the highest and lowest deviant points. When the S/N of the sub-array observations is sufficiently high, we adopted the intrinsic instrumental uncertainty, 0.01 and 0.007 magnitude at 3.6 and 4.5 $\micron$, respectively \citep{reb14}, as the error of the sub-array photometry. All other targets were observed in the full array mode.

The photometry of all sources is listed in Table~\ref{spitzer_phot}. The nominal errors obtained directly from the S/N on the images are mostly slightly better than 1\%, generally consistent with the expected instrumental performance\footnote{IRAC Instrument Handbook v2.0.3, http://irsa.ipac.caltech.edu/data/SPITZER/docs/\\irac/iracinstrumenthandbook/17/.}. As an independent evaluation, for P1121 (2MASS J07354269-1450422) we measured some field stars with similar brightnesses and found photometric stability of about 1\%. We do not have field stars in the same image with comparable brightness for other targets.

To separate the stellar and disk contributions to the total fluxes, we used Kurucz atmosphere models \citep[ATLAS9][]{cas04} to fit the available optical and near-infrared observations. In the optical, for better accuracy only the Hipparcos/Tycho catalogs \citep{per97,hog00}, photometric surveys \citep{dro06,ofe08}, and dedicated photometry \citep[e.g.,][]{pri03} were used. 2MASS photometry \citep{skr06} was adopted for the near-infrared. Given the range of spectral types of our sample stars, the modeled photospheric fluxes are not sensitive to the assumed stellar gravity or metallicity within reasonable ranges of main sequence stars ($<5\%$ difference, which should have no significant effect on our later analysis). For simplicity, we fixed the stellar gravity and metallicity to be solar for all our targets, and used $\chi^2$ statistics to determine the best-fit parameter for stellar effective temperature by fitting all available optical and near-infrared photometry. For more distant objects ($>$100 pc), corrections for interstellar extinction ($A_V$ estimated from $E(B-V)$ colors) were also applied before the $\chi^2$ fitting. With the adopted distances (either from Hipparcos parallax or cluster location), the integrated stellar luminosities were also checked to be consistent with the expected values for stars in the main sequence. The expected photospheric fluxes in the IRAC 3.6 and 4.5 $\micron$ bands were computed using the IDL code, {\it spitzer\_synthphot}\footnote{http://irsa.ipac.caltech.edu/data/SPITZER/docs/\\dataanalysistools/cookbook/14/}, provided by the SSC. The best-fit model and relevant stellar parameters along with the adopted stellar photospheric values are given in Table~\ref{spitzer}. The photospheric contributions at the two IRAC wavebands are subtracted from the observed total fluxes to obtain the fluxes of the debris disks.

\subsection{Analysis Methods}

We apply three discriminants to the infrared time-series data to identify debris disk variability. A disk is considered variable if it passes at least one of them. The first one is the Stetson index \citep{ste96}, which assumes that real variations will cause correlated light curves between paired observations (typically pairs of near simultaneous photometry at two wavelengths). The index, $S$, is defined as
\begin{equation}
S = \frac{ \displaystyle\sum_{k=1}^{N} g_k \, \textrm{sgn}(P_k) \sqrt{\left| P_k \right|} }{\displaystyle\sum_{k=1}^{N} g_k}
\end{equation}
where $g_k$ is the weight of the $k$-th epoch, $P_k \equiv \delta_{\lambda_1,k} \delta_{\lambda_2,k}$ for the $k$-th observations at wavelengths $\lambda_1$ and $\lambda_2$, respectively, where
\begin{equation}
\delta_k = \sqrt{ \frac{N}{N-1} } \frac{a_k - \langle a \rangle}{\sigma_{a_k}}
\end{equation}
where $N$ is the total number of measurements in the time-series, $a_k$ is the $k$-th data point up to $k = N$, $\sigma_{a_k}$ is its error, and $\langle a \rangle$ is the mean of all $a_k$. An index of $S=0$ indicates no correlation between the two sets of light curves. Positive and negative indices suggest correlation and anti-correlation, respectively. The larger the absolute value is, the stronger the correlation is. Since noise is statistically unlikely to be correlated, a strong correlation between data at different wavelengths is usually a good indicator of bona fide variations. In this work, since all observations of each individual source are conducted with exactly the same AOR design, we opt to compute the Stetson index with unweighted data, i.e., $g_k = 1$ for all.

Previous works have used different thresholds of Stetson index for IRAC 3.6 and 4.5 $\micron$ to identify variability. For example, \citet{fla13} adopted $S=0.45$ for IC 348 as confirmed with $\chi^2$; \citet{cod14} found $S=0.21$ for 3-$\sigma$ confidence in NGC 2264 based on the distribution of the indices of field stars; \citet{reb14} adopted a much more rigorous $S = 0.9$ for $\rho$ Oph, but that corresponds to 6-$\sigma$ for the fitted Gaussian distribution. In this work, the fields of our sample stars are much less nebulous than those in IC 348 or $\rho$ Oph, permitting a less conservative threshold than is appropriate for them. We adopt a threshold of $S = 0.5$.

The second discriminant is a $\chi^2$ test for the variance of the observations, given by
\begin{equation}\label{chi-square}
\chi_{red}^2 = \frac{1}{N-1} \displaystyle\sum_{k=1}^{N} \left( \frac{a_k - \langle a \rangle}{\sigma_{a_k}} \right)^2
\end{equation}
where $\sigma_{a_k}$ is the nominal uncertainty of the $k$-th photometry, directly measured from the signal-to-noise ratio of the corresponding image. The number of degrees of freedom is $(N - 1)$ because the only fitted parameter is the mean value, $\langle a \rangle$. Given the complexity of the BCD-based photometry and non-Gaussian instrumental fluctuations, the estimated errors of individual measurements ($\sigma_{a_k}$) are not expected to be very accurate. To accommodate the additional uncertainty, higher values, like 3 \citep{fla13} or 5 \citep{reb14}, are usually adopted as the threshold for confident detection of variations. To be conservative, here we adopt the more rigorous $\chi_{red}^2 \geq 5$.

Both the Stetson index and the $\chi_{red}^2$ test ignore the time-resolved nature of light curve monitoring. When they both fail to reveal significant variations, as a supplement, we fit a linear trend to the unweighted data to see if the slope is significantly different than 0. Such a fit utilizes the time-resolved information and should be sensitive to a monotonic trend of flux. The values of the variability discriminants of each source are listed in Table~\ref{variability}.

In addition to the identification of the variability of disk emission, we also use $\chi^2$ to look for color variations of the disks during our observations. Because obtaining the disk color requires the subtraction of stellar photospheric flux, which is likely a source of systematic offset, we do the $\chi^2$ test for the color of the entire system (star + disk) as well as for the disk only. Considering the large uncertainty of the errors on disk color, we conservatively considered disk color to be variable only if either the disk or the entire system had $\chi_{red}^2 \geq 6$.

If a disk is found to be variable, we will also analyze both the long-term evolution trend of its flux and any periodicity on top of that. We computed the periodograms of all time-series data, including those in the optical (see appendix), to look for periodicities with the SigSpec algorithm \citep{ree07}, which is an extension to the Lomb-Scargle periodogram and takes both the amplitude and phase of the Fourier transform into account. To reliably identify periods, we adopted a threshold of spectral significance, or $sig$, of 5 for ``significant'' peaks, which corresponds to an amplitude S/N of $\sim$3.0.

When we have data with high sampling frequencies, the results from the SigSpec algorithm are double checked with a second algorithm. For this purpose, we use the Plavchan periodogram \citep{pla08}, which is a binless phase dispersion minimization algorithm based on dynamical priors generated from the data and has no pre-assumed wave function. This algorithm can detect periodic signals of any arbitrary shape. Our data for most sources are not dense and long enough to reveal fine characteristics in the time domain that demand analysis by multiple algorithms. In this work, the second method is only used for BD+20 307, for which we have high cadence data in two visibility windows.

For sparse infrared data where significant periodogram analysis is unachievable, we also turn to the discrete auto-correlation function ($ACF$) to look for a characteristic timescale.
\begin{equation}
ACF(\tau) = \frac{1}{(N - \tau) \sigma_a^2} \displaystyle\sum^{N - \tau}_{k=1} \left(a_k - \langle a \rangle \right) \left(a_{k+\tau} - \langle a \rangle \right)
\end{equation}
where $\tau$ is the number of time steps, and $\sigma_a$ is the standard deviation of all $a_k$. By definition, $ACF (\tau = 0) = 1$. The characteristic timescale is defined at the position of the first local maximum. $ACF$ works for both periodic and aperiodic time-series; the characteristic timescale does not necessarily mean periodicity or quasi-periodicity.

\section{Results and Discussions of Individual Stars}

\subsection{ID8}

ID8 is a G6V star in the 35-Myr-old \citep{sod14} open cluster NGC 2547 \citep{gor07}. The time domain observations and analysis of its disk in 2012 and 2013 are reported in an earlier paper \citep{men14}, and provide the prototype for the analysis of other debris disks in this work. Here we summarize the conclusions from the study of this star.

ID8 has an extreme debris disk with fractional luminosity of $3 \times 10^{-2}$ \citep{olo12}. The debris disk is one of the first found to vary on yearly timescales based on multi-epoch MIPS 24 $\micron$ observations; comparisons between IRAC observations, IRS synthetic photometry, and {\it WISE} data suggested that the variations could also be seen at shorter wavelengths \citep{men12}.

At an ecliptic latitude of $-66 \arcdeg$, ID8 is well positioned for a continuous {\it Spitzer} visibility window of $\sim$221 days every year. Benefiting from this, \citet{men14} monitored the disk with IRAC for about 3 months in mid-2012, and then in the new visibility window starting from January 2013. In the optical wavebands, the star appeared to be quiescent throughout the visibility window in 2013, so stellar variability can be excluded as the driving force of the disk variations. Meanwhile in the infrared, though with significant variations, the disk flux was generally flat until mid-2012. But at the beginning of 2013, it had brightened by $\sim50\%$ at both 3.6 and 4.5 $\micron$. The flux then followed a general decay with a timescale of 370 days at both wavelengths, plus quasi-periodicity of 25 and 34 days. Color variations were also detected, and tend to suggest a combination of changing dust temperature and changing dust emitting area, or in area alone, as the direct cause of the disk flux variations.

The two years' evolution of the ID8 disk flux indicates a real-time large impact not long before the observations in 2013 \citep{men14}. In the model for this event, a hypervelocity impact between planetary embryos, or at least between large planetesimals, produced a silica-rich vapor plume, from which silicate spherules condense to form an optically thick dust cloud. Collisions among these spherules broke them down into fine dust. The depletion of large feeding grains and the loss of fine particles with sizes below the radiation blowout limit in the stellar radiation field caused continuous attenuation of the disk, corresponding to the decay of the disk flux over a year. As the optically thick dust cloud revolved around the star on a moderately eccentric orbit, the recurring geometry produced quasi-periodic modulations of disk flux as we saw from the Earth.

\subsection{P1121}

P1121, cataloged as 2MASS J07354269-1450422, is a member of the 80-Myr-old \citep{roj97} open cluster M47. P1121 was first noticed in 2004 for its extraordinary 24 $\micron$ excess \citep{gor04}. The optical spectrum shows a spectral type of F9 V \citep{gor04}. We have estimated the fractional luminosity of the disk to be $\sim 2 \times 10^{-2}$.

\subsubsection{Disk SED}

\begin{figure}
\epsscale{1}
\plotone{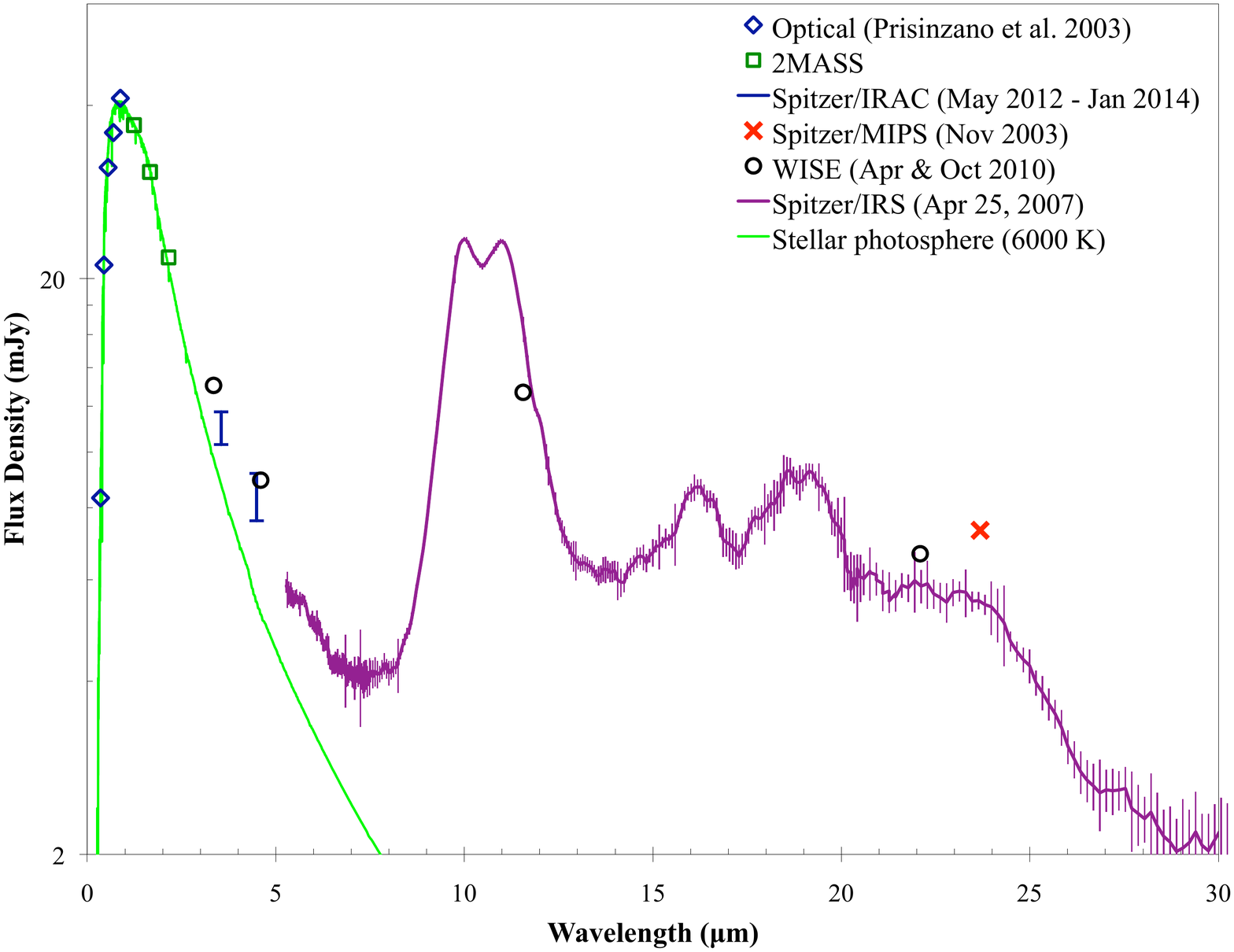}
\caption{SED of P1121. The vertical bars with caps at 3.6 and 4.5 $\micron$ represent the full range of the disk flux observed at each waveband in this work, from 2012 to early 2014.
\label{P1121_SED}}
\end{figure}

The mid-infrared spectrum of the P1121 disk (Figure~\ref{P1121_SED}), observed on April 25, 2007, is highly analogous to that of ID8 \citep{mor14}. A detailed mineralogical model will be presented by N. Gorlova et al. (2015, in preparation). According to their results, the P1121 disk is dominated by sub-$\micron$-sized amorphous and crystalline silicate dust, and is very similar to that of ID8 in composition regardless of some minor differences. Conventionally, the dust sizes within a debris disk follow a power law distribution. Making this assumption, the best fit power law index for the P1121 disk is $-4.0$. This is much steeper than typical values near $-3.65$ \citep{doh69,gas12}, but close to the fragment size distribution after laboratory hypervelocity impacts \citep[$-4.4 \pm 0.8$,][]{tak11}, and those found around HD 172555 \citep[$-3.95 \pm 0.10$,][]{joh12b} and ID8 \citep[$-4.0 \pm 0.2$,][]{olo12} where recent large impacts are either suspected or confirmed. 

\subsubsection{Infrared Light Curve}

\begin{figure}
\epsscale{1}
\plotone{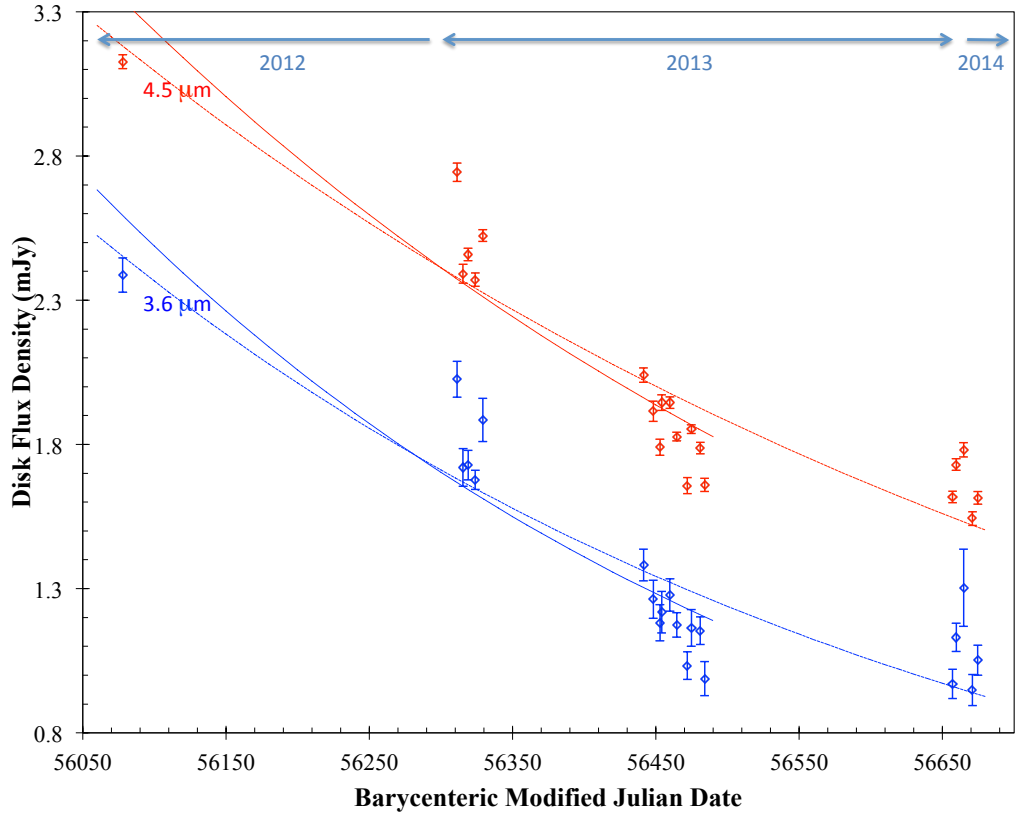}
\caption{Infrared light curve of the P1121 disk. Overlaid are the fits to the data assuming they follow an exponential decay. The dashed and solid lines represent fit 1 (including all the data) and fit 2 (excluding the 2014 data), respectively.
\label{P1121}}
\end{figure}

In the time domain, the infrared behavior of the disk resembled the characteristics we saw in the ID8 disk \citep{men14}. As shown in Figure~\ref{P1121}, since 2012 the disk flux has been on a downward trend, on top of which are hints of significant variations on some much shorter timescale. Following the analysis of ID8 \citep{men14}, we empirically approximate the first order trend of the excess evolution with an exponential decay. Fitting all the data using $\chi^2$ minimization finds a timescale of $620 \pm 70$ days at 3.6 $\micron$, and $800 \pm 70$ days at 4.5 $\micron$ (fit 1). The errors of the decay timescales are from the goodness of the fits based on nominal photometric errors of the data. We notice that the disk flux in 2014 was not much lower than in late 2013. A better fit could be achieved without the 2014 data, as would be appropriate if there was a modest new injection of additional dust in that year, in which case the decay timescales would be $530 \pm 50$ and $680 \pm 70$ days at 3.6 and 4.5 $\micron$, respectively (fit 2).

\begin{figure}
\epsscale{1}
\plotone{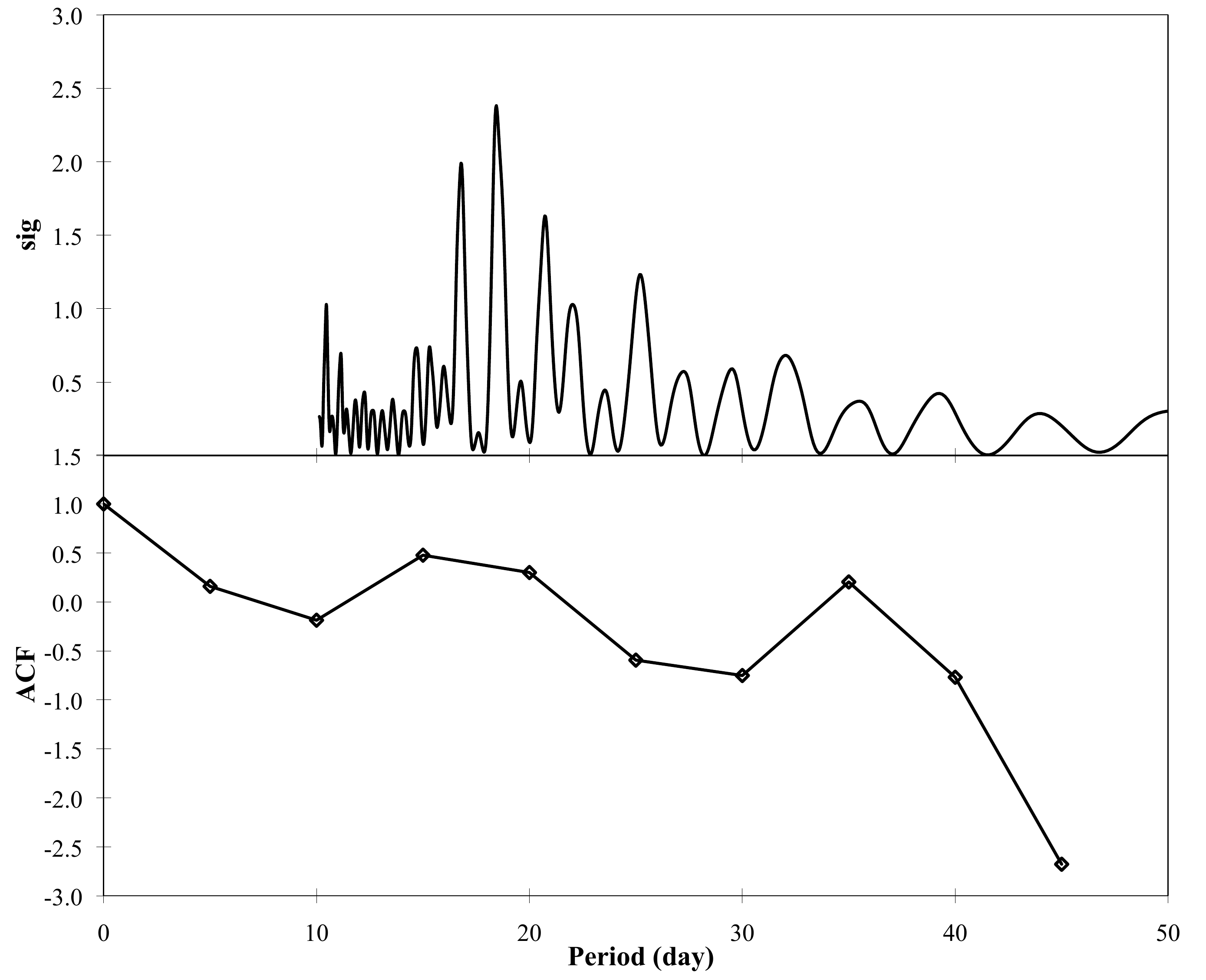}
\caption{Analysis of the detrended 4.5 $\micron$ data of the P1121 disk. Upper: SigSpec periodogram. The strongest signal is at $\sim$18 days with $sig = 2.65$. Lower: $ACF$ with a time step of 5 days. The first local maximum between 15 and 20 days is the characteristic timescale, corroborated by the second local maximum around 35 days.\label{P1121_period}}
\end{figure}

As in \citet{men14}, for periodogram analysis we focus on the 4.5 $\micron$ data where the disk is measured at a higher S/N. After the best-fit exponential decay function is subtracted, the ``detrended'' disk flux time-series shows no significant period. The strongest signal in the periodogram has $sig = 2.65$, at $\sim$18 days with an amplitude of 0.134 mJy (Figure~\ref{P1121_period}). However, the lack of a well determined signal is likely because our limited time coverage and sampling frequency were too sparse to identify short periods at high significance.

Although each {\it Spitzer} visibility window of P1121 lasts $\sim$49 days, only the second one in 2013 was fully covered because of observation scheduling issues. The observation sampling was not equidistant even in this visibility window. So, we have to interpolate the original time-series to a time grid from BMJD 56440 to 56485, with steps of 5 days. The final $ACF$ of the P1121 disk is plotted in Figure~\ref{P1121_period}. The characteristic timescale is defined at the position of the first local maximum, in this case between 15 and 20 days. The second local maximum around 35 days is close to twice the period of the first maximum, corroborating its reality.

Interestingly, the $ACF$ characteristic timescale is in agreement with the most plausible (although not ``significant'') period in the periodogram. An examination of this timescale with the data in other visibility windows does not find obvious inconsistency. Hence, the modulation over 15-20 days of the P1121 disk might be valid, pending verification and more accurate determination by more observations.


\subsubsection{Disk Color}

As in the case of ID8 \citep{men14}, the nominal uncertainties of the photometric measurements, including the star and the disk, are $\lesssim1\%$ at both wavelengths. However, since the star contributes a considerable portion ($>$75\%) of the total flux at both 3.6 and 4.5 $\micron$, after subtracting the stellar fluxes as constants, the relative errors with respect to the disk flux are 1-5\%, and can be $>10\%$ at 3.6 $\micron$ when the disk is faint. In addition, errors of the adopted stellar flux would introduce a systematic bias to the disk flux and color, which can hardly be identified or corrected. The relative disk flux and color variations are immune to such effects. But the absolute color index and color temperature of the disk may be biased, and thus their nominal errors could be misleading. Therefore, we only consider the average and range of their absolute values without quoting errors, keeping in mind they may be subject to some systematic offset.

\begin{figure}
\epsscale{1}
\plotone{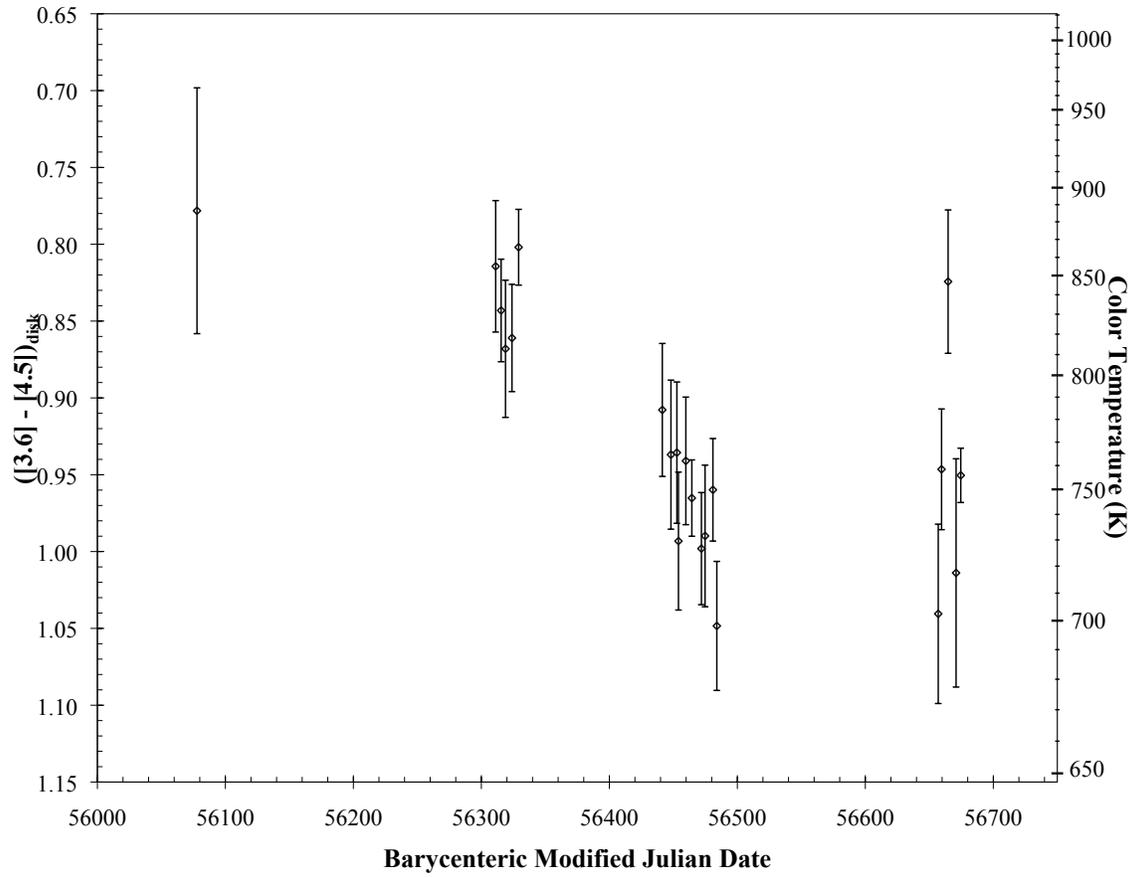}
\caption{Evolution of the color index and temperature of the P1121 disk.\label{P1121_Tt}}
\end{figure}

With observations at two wavelengths, we find the $\chi_{red}^2$ value of the color index, $[3.6]-[4.5]$, is 7.7 for the entire P1121 system and 3.8 for the disk flux, indicating significant color variations. After subtracting the stellar photospheric emission, the average color of the disk is $[3.6] - [4.5] = 0.92$, corresponding to a temperature of 770 K. The range of the temperature variations, subject to the systematic error of the adopted photospheric flux, is from 700 to 890 K. Even with our very conservative error bars, the color index and temperature evolution appear to be decreasing over time (Figure~\ref{P1121_Tt}).



\subsection{HD 15407A}

HD 15407 is a binary of two main-sequence dwarfs of F5V and K2V, separated by 21\arcsec, or 1200 AU at a distance of 55 pc. It may be a member of the AB Dor Moving Group, with an age determined by lithium absorption of 80 Myr \citep{mel10}. The excess of component A, $f_d \sim 6 \times 10^{-3}$ \citep{mel10}, was first observed by IRAS in 1983 \citep{oud92}.

\subsubsection{Disk SED}

\begin{figure}
\epsscale{1}
\plotone{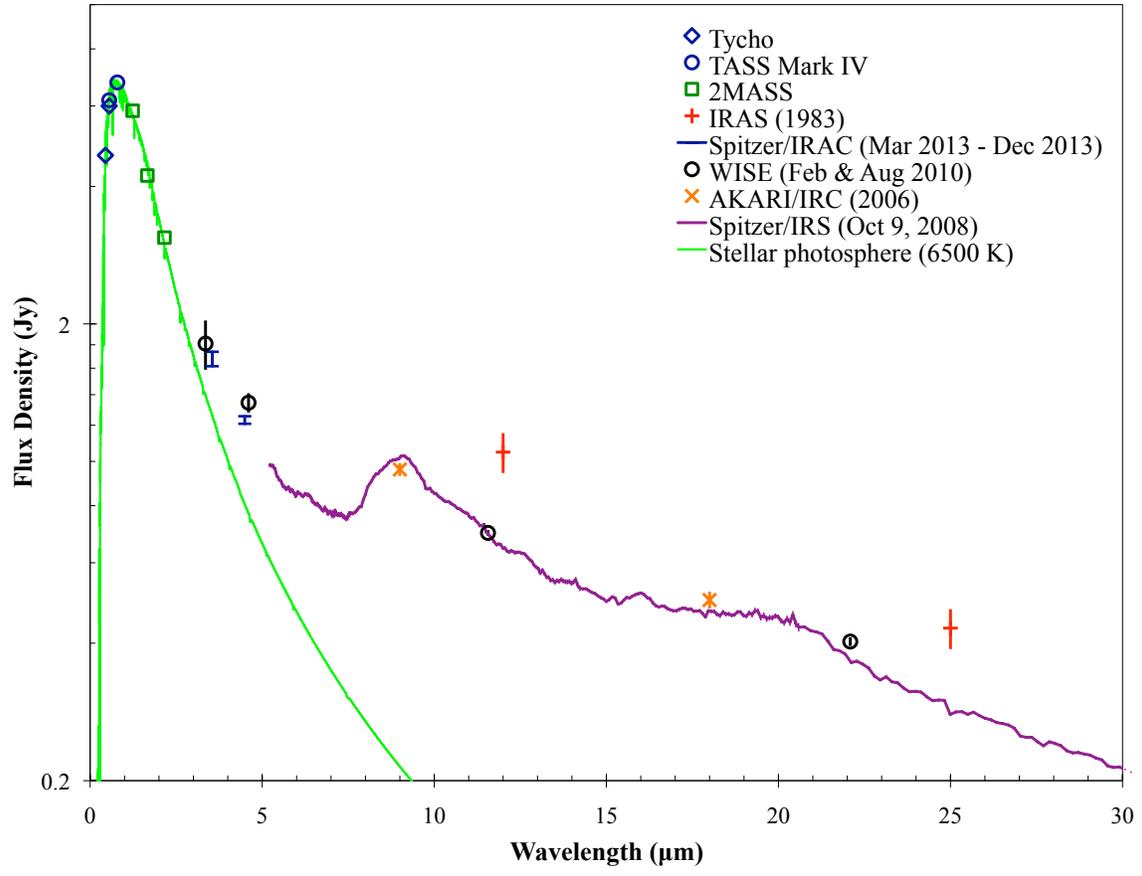}
\caption{SED of HD 15407A from the optical to mid-infrared wavelengths. The vertical bars with caps at 3.6 and 4.5 $\micron$ represent the full range of the disk flux observed in this work. See \citet{fuj12b} for data at longer wavelengths up to far-infrared.\label{HD15407A_SED}}
\end{figure}

The only mid-infrared spectroscopic observation, taken on October 9, 2008 by {\it Spitzer}/IRS and shown in Figure~\ref{HD15407A_SED}, has been modeled by different groups \citep{mel10,fuj12a,olo12}. These models basically agree on a disk with fine dust particles of both silica and amorphous silicate, plus blackbody-like large dust grains slightly hotter than 500 K. In addition, the {\it WISE} data revealed an anomalous excess at 3.4 and 4.6 $\micron$, which might be attributed to hotter dust with temperature $\gtrsim$1000 K, or to the SiO gas emission band from recent collisions \citep{fuj12b}. The ranges of our 3.6 and 4.5 $\micron$ photometry are generally consistent with the {\it WISE} measurements in 2010, as shown in Figure~\ref{HD15407A_SED}.

\subsubsection{Infrared Light Curve and Color}

\begin{figure}
\epsscale{1}
\plotone{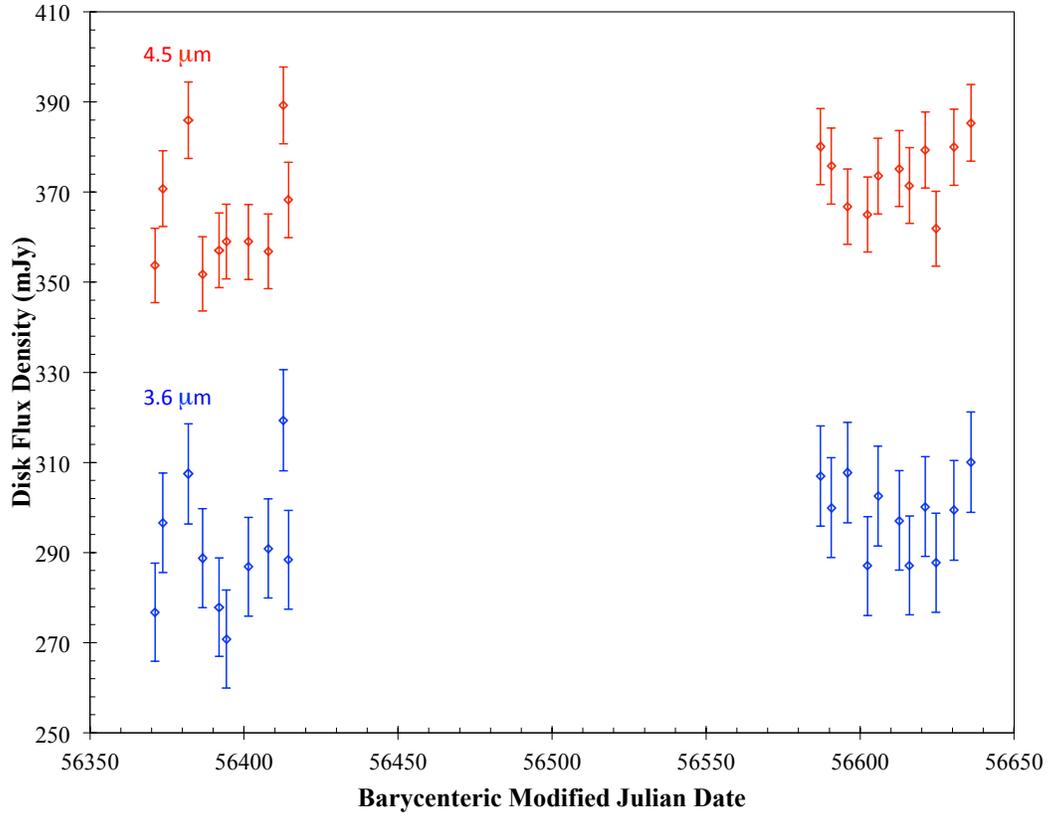}
\caption{Infrared light curve of the HD 15407A disk in 2013. Synchronous variations between the two wavebands are apparent with no clear monotonic trend.\label{HD15407A}}
\end{figure}

The time series is plotted in Figure~\ref{HD15407A}. Mutually synchronous variations are evident in the two wavebands with a Stetson index of 0.81, above the commonly adopted thresholds for variability. However, unlike ID8 or P1121, the disk around HD 15407A showed neither a decay trend over our total time baseline of $\sim$265 days, nor any signs of periodicity in the periodogram and $ACF$.

Considering both wavebands, the disk of HD 15047A has a mean color of $[3.6] - [4.5] = 0.73$. Assuming a disk of blackbody, this gives a temperature of 930 K. A $\chi^2$ test based on the nominal photometric errors suggests that the color consistency of the disk and of the entire system were both $\chi_{red}^2 = 0.2$. That is, within the likely errors, there is no firm detection of color variations. We found that the result is robust and insensitive to the adopted photospheric fluxes. Compared with the disks of ID8 and P1121, both the flux and color of the HD 15407A disk were more stable throughout the observations. 


\subsection{HD 23514AB}

A member of the $\sim$120-Myr-old Pleiades cluster, HD 23514 is one of the prototypes of extreme debris disks that were found variable over yearly timescales at 24 $\micron$, at which wavelength the disk flux was observed to have decreased by nearly 10\% in about 200 days in 2007 \citep{men12}. With a spectral type of F5V, the star has a dusty disk with high excess ($f_d \sim 2 \times 10^{-2}$) in its terrestrial planet zone \citep{rhe08}. The star was thought to be single until a $\sim$M8 brown dwarf companion was recently discovered at a separation of 2.64\arcsec\ \citep{rod12,yam13}. This separation, equivalent to 2.2 native pixels compared to the photometric aperture radius of 3 pixels, is too small to be effectively resolved by IRAC. Our direct photometric measurements will have included emission from both stars.

To obtain the expected companion flux at the observed wavelengths, brown dwarf spectrophotometry is adopted from the BT-Settl models \citep{all11}, and is integrated over the transmission curve of the IRAC wavebands. The expected flux is then corrected by an aperture correction factor to account for the offset of the photometric aperture from the companion position. The aperture correction factor is 0.861 at 3.6 $\micron$ and 0.845 at 4.5 $\micron$ based on the offset distance and the instrumental PRF of the IRAC detector. Finally, the estimated companion contribution is subtracted from both wavebands in addition to the removal of the flux of the primary star. The expected flux of HD 23514B is given in Table~\ref{spitzer}, and is only $< 0.3\%$ of the flux from HD 23514A. So it has a negligible effect on our photometry.

\begin{figure}
\epsscale{1}
\plotone{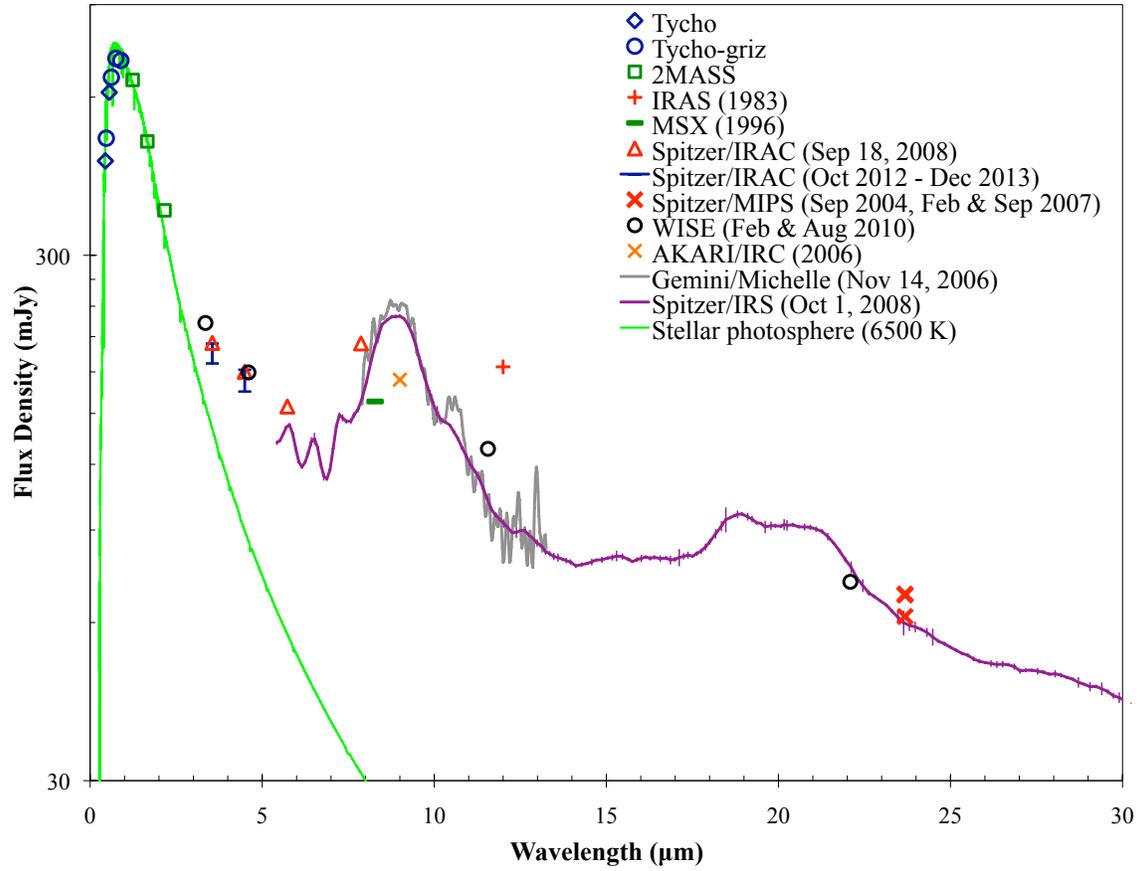}
\caption{SED of HD 23514. The vertical bars with caps at 3.6 and 4.5 $\micron$ represent the full range of the disk flux observed in this work. For clarity, the errors of the Gemini/Michelle observation are not plotted. The Midcourse Space Experiment (MSX) astrometry is off by 12.7", but is likely associated with HD 23514 \citep{rhe08}.\label{HD23514_SED}}
\end{figure}

The mid-infrared SED of HD 23514, as in Figure~\ref{HD23514_SED}, is strongly peaked at 9 $\micron$ \citep{rhe08}, indicative of a silica-rich disk, which differs from most other debris disks that are dominated by silicates \citep{mor14}.

\subsubsection{Infrared Light Curve}

\begin{figure}
\epsscale{1}
\plotone{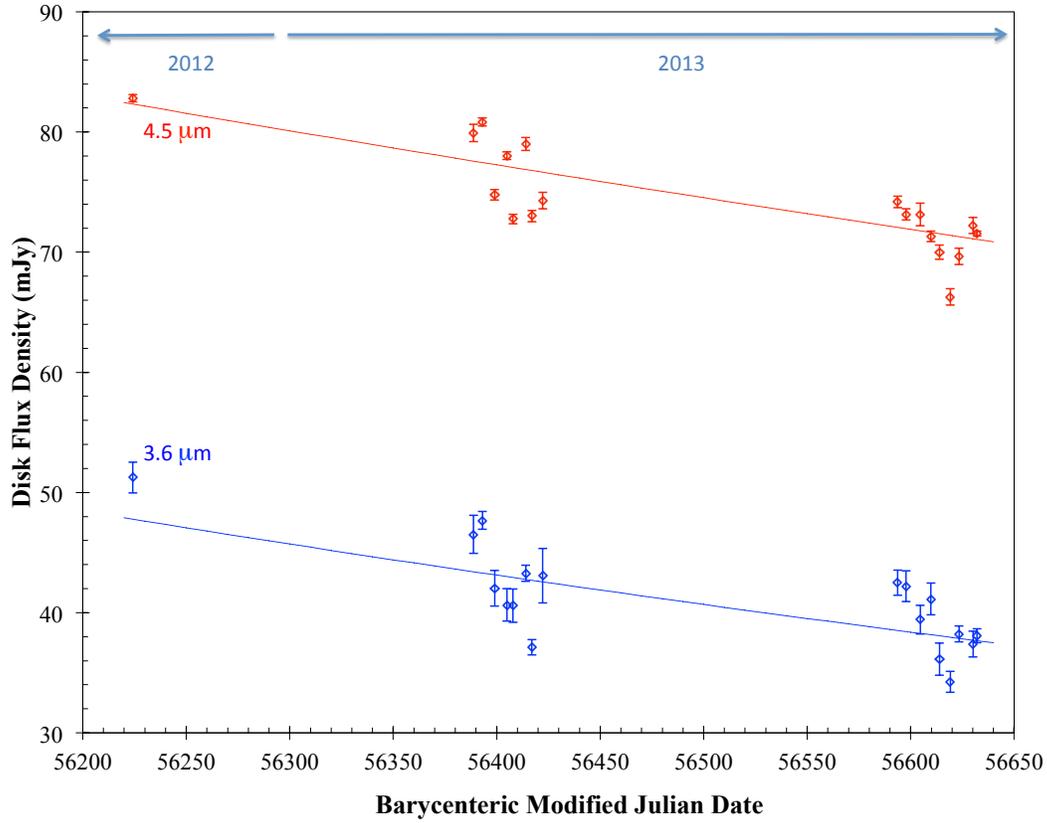}
\caption{Infrared light curve of the HD 23514A disk. Overlaid lines are the best-fit exponential decay to the data in both wavebands. However, because the fitted timescales are much longer than the time baseline of observations, the assumption of a monotonic decay may be problematic and the fits may be purely phenomenological.\label{HD23514}}
\end{figure}

The light curve of HD 23514 (Figure~\ref{HD23514}) appears to have a decay trend. An exponential fit finds that the decay timescale is $1720 \pm 140$ days at 3.6 $\micron$, and $2780 \pm 70$ days at 4.5 $\micron$. Such decay timescales of the disk flux are derived based on large extrapolations from a time baseline of only 408 days in 2012 and 2013. It is equally likely that the trend is essentially part of a period or a segment of some random evolution over longer timescales, rather than the kind of monotonic decay seen in the disks around ID8 \citep{men14} and P1121.

After subtracting the best-fit exponential trend, the infrared time-series shows no significant period at either waveband. Data in the first 38-day-long visibility window of 2013 appeared to fluctuate as in the case of P1121. But $\sim$172 days later, in the second visibility window the disk flux followed a smoother evolution with a dip at BMJD 56619.2. This contradicts the plausible characteristic timescale in the first visibility window. As a result, we are unable to determine any periodicity of the HD 23514 disk.

\subsubsection{Disk Color}

The mean disk color of HD 23514 during our observations is $[3.6] - [4.5] = 0.96$, equivalent to a blackbody temperature of 750 K. This matches well with the disk temperature derived from the mid-infrared SED \citep{rhe08}. A $\chi^2$ test indicates that the color consistency of the disk and of the whole system were $\chi_{red}^2 = 5.4$ and $3.3$, respectively. The values are close to, but fall short of our threshold of 6. Accordingly, the observed color variations are not considered significant. But future monitoring of this system may be helpful. 


Unlike in the case of HD 15407A, the SED of HD 23514 does not show the anomalous excess at the 3-5 $\micron$ region that might be attributed to emission by very hot dust or SiO gas. The observed excess from HD 23514 is likely from the thermal emission of solid warm dust particles in the disk.

\subsection{HD 145263}

HD 145263 is a F0V star in the $\sim$11-Myr-old Upper Sco Association \citep{pec12}, and the earliest type star in our sample of potentially variable debris disks. The star is likely to be single, as an imaging search for multiplicity has returned a null result between 0.1\arcsec\ and 5.0\arcsec\ down to planetary mass \citep{jan13}.

\subsubsection{Disk SED}

\begin{figure}
\epsscale{1}
\plotone{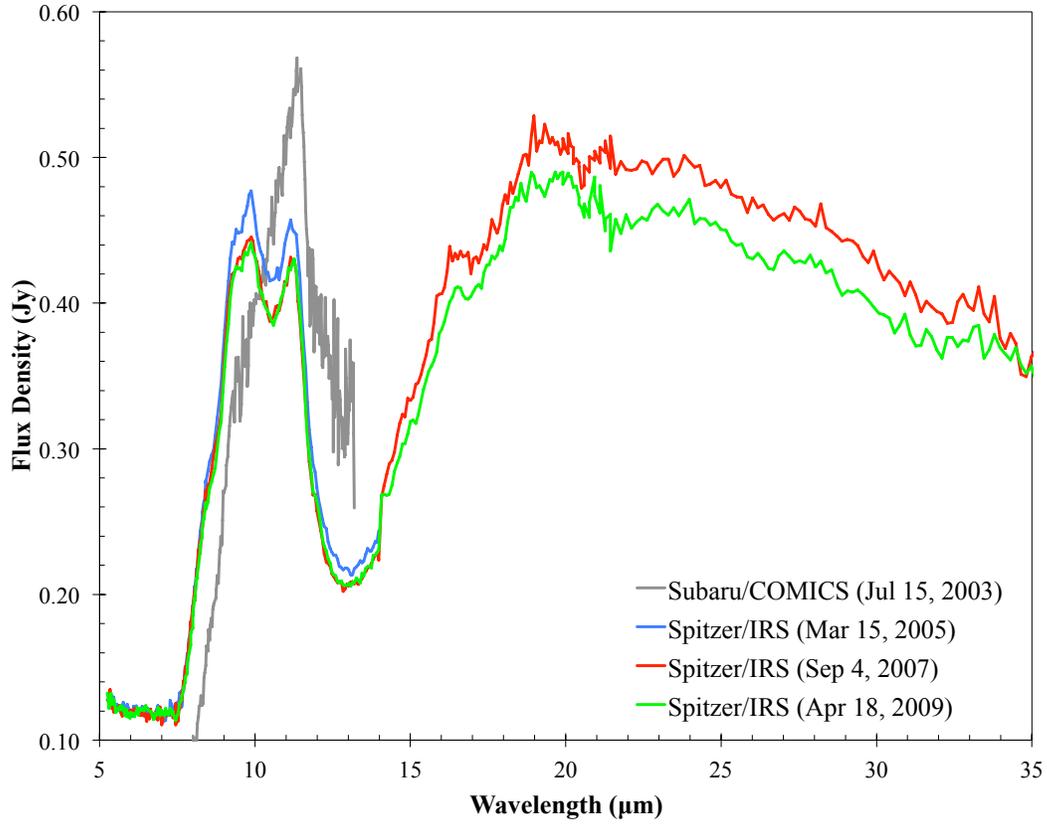}
\caption{Comparison of the {\it Spitzer}/IRS SEDs of HD 145263 at different epochs. For clarity, error bars are not plotted. The relative errors in flux density are mostly between $2\%$ and $10\%$ at all epochs, with an average of $5.2\%$. The ground-based Subaru/COMICS observation in 2003 \citep{hon04} is overplotted for reference without error bars.\label{HD145263_multiepoch}}
\end{figure}

HD 145263 is the only star that has been spectroscopically observed by {\it Spitzer}/IRS over multiple epochs, in March 2005, September 2007, and April 2009, respectively. We ensured homogeneous data reduction by adopting the optimally extracted CASIS spectra \citep{leb11} for all the observations. In addition, the system was observed with Subaru/COMICS at 8 - 13 $\micron$ in July 2003 \citep{hon04}. The results are compared in Figure~\ref{HD145263_multiepoch}. In the 2005 IRS observation, the longer wavelength part of the spectrum was taken in LH mode and had no CASIS reduction. Our reduction of the data shows that the disk was likely brighter than in 2007 and 2009 by $\lesssim$10\% over most wavelengths. However, the 2005 data had no accompanied sky observation, and require a scaling factor to make the transition between SH and LH smooth. The determination of this factor is somewhat arbitrary, depending on the reference photometry adopted and how mineralogical features in the transition region are handled. For the most homogenous comparison between the IRS data, we do not show the long wavelength part of the 2005 spectrum.

We find that the Subaru/COMICS spectrum in 2003 shows very different mineralogy from all IRS spectra at later epochs. Comparatively, the IRS spectrum in 2005 reveals much stronger emission at 8 - 9 $\micron$ that is likely from silica \citep{spi61}, but weaker 11.4 $\micron$ feature from crystalline forsterite \citep{hon03,hon04}. The dramatic changes may imply a major collision event between the two epochs. From 2005 to 2009, each SED has slightly decreased flux from the previous epoch almost over the entire spectrum with no more significant changes in the spectral structure that corresponds to the chemical mixes of the disk (C. Lisse, private communication). The largest differences of the IRS observed flux among different epochs are $\sim$8\%, occurring near the silica/silicate emission features around 10 and 20 $\micron$.

\subsubsection{Infrared Variability}

\begin{figure}
\epsscale{1}
\plotone{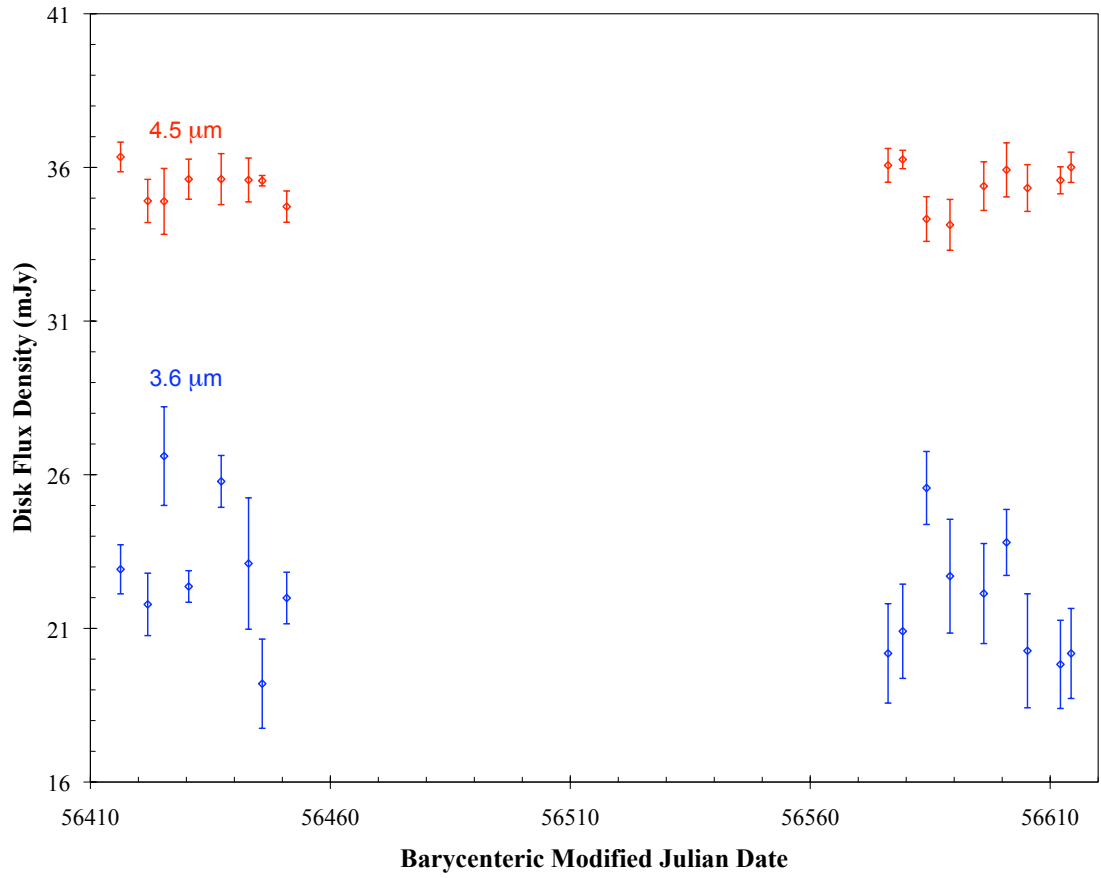}
\caption{Infrared light curve of the HD 145263 disk in 2013. Analysis reveals no significant variations in 2013.\label{HD145263}}
\end{figure}

The time series of the disk emission of HD 145263 is shown in Figure~\ref{HD145263}. The photometric uncertainties at both wavebands are $\lesssim$1\%. But since 90\% and 78\% of the observed total fluxes at 3.6 and 4.5 $\micron$ are from the star, the relative errors of the disk flux measurements are 2-9\% at 3.6 $\micron$ and 1-3\% at 4.5 $\micron$.

Based on the data in both wavebands in 2013, HD 145263 had a Stetson index of $S = -0.28$. The absolute value of the index does not compare favorably against commonly adopted thresholds. Unfortunately, our field of view for HD 145263 does not contain enough field stars to warrant an independent evaluation for the threshold, while our other debris disk samples are all easily identified as variables and have little value as references for this case.

When it comes to the $\chi^2$ test, the time-series of HD 145263 yielded $\chi_{red}^2 = 3.0$ at 3.6 $\micron$ and $\chi_{red}^2 = 1.4$ at 4.5 $\micron$. These values also fall short, or look ambiguous at most, compared to commonly used thresholds of 3 or 5.

Finally, we find that the slope of the flux of HD 145263 was $-3.2 \pm 2.3$ mJy per year at 3.6 $\micron$ and $0.1 \pm 0.7$ mJy per year at 4.5 $\micron$ over our total time baseline of  $\sim$198 days in 2013. Neither was significantly different from 0.

In conclusion, we did not detect any significant disk variation of HD 145263 over the IRAC monitoring in 2013. Given the short time covered by our observations, the lack of variations in our observations is not necessarily inconsistent with the long-term decay suggested by the decreased spectral flux from 2005 to 2009.

The color of the disk is found to be $[3.6] - [4.5] = 0.99$, corresponding to a blackbody temperature of 730 K.

\subsection{BD+20 307AB}

\begin{figure}
\epsscale{1}
\plotone{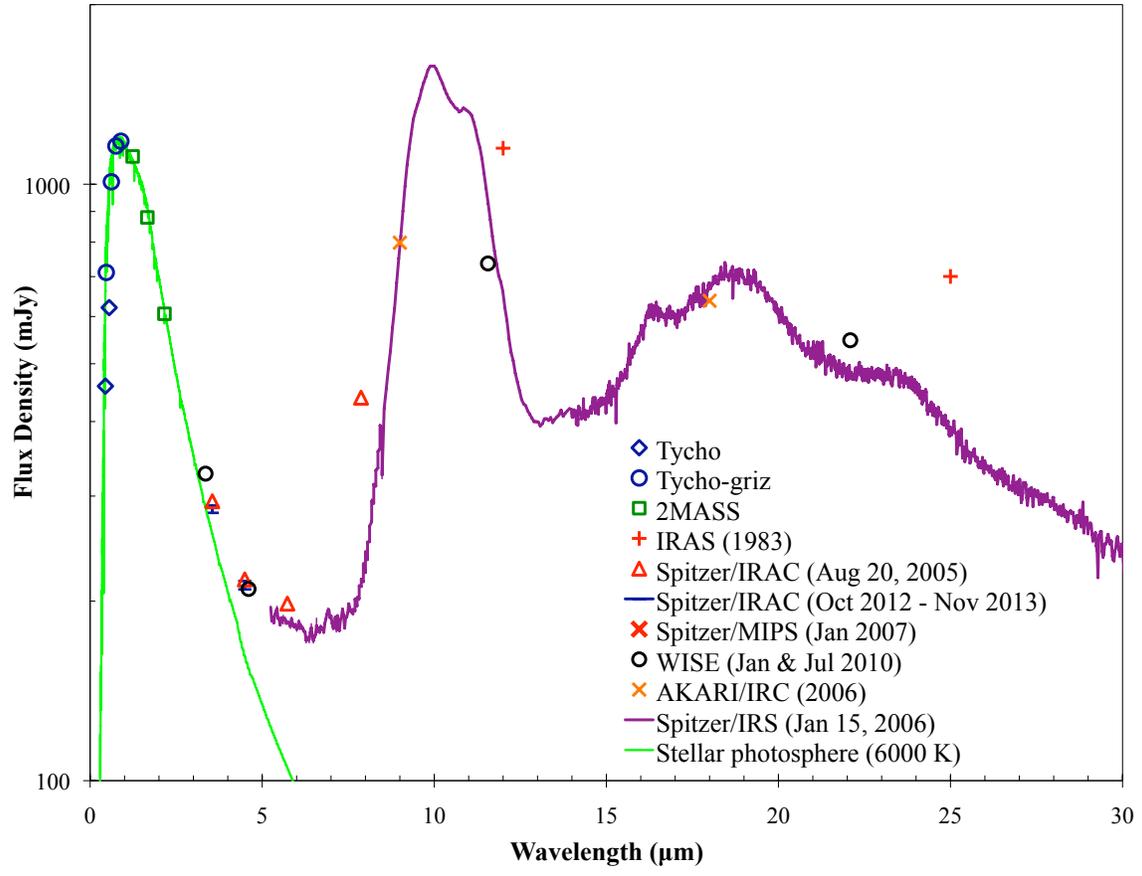}
\caption{SED of BD+20 307. The vertical bars with caps at 3.6 and 4.5 $\micron$ represent the full range of the disk flux observed in this work, and are directly comparable with the one-epoch data point from cyrogenic {\it Spitzer}/IRAC observations.\label{BD20307_SED}}
\end{figure}

The high fractional luminosity $f_d \sim 4 \times 10^{-2}$ of BD+20 307 (Figure~\ref{BD20307_SED}) was first observed by IRAS \citep{oud92}. Early in the {\it Spitzer} era, the star was thought to be single and in the age range of terrestrial planet formation \citep{son05}. But high resolution spectra revealed that it is a spectroscopic binary of two stars of nearly identical G0V spectral type with a mutual orbital period of 3.42 days, and the revised age of the system is beyond 1 Gyr \citep{wei08,zuc08}. Both stars contribute to the flux we observed in the monitoring. But since they have nearly identical temperatures and spectral types \citep{zuc08}, in this work we fit and subtract their photospheric contribution with a single stellar model. No other component is found with adaptive optics and four seasons of radial velocity measurements \citep{fek12}. Despite the high infrared excess, a SED model found no trace of a cold disk component \citep{wei11}.  The disk is known to be variable from {\it WISE} observations in 2010 before our monitoring \citep{men12}.

\begin{figure}
\epsscale{1}
\plotone{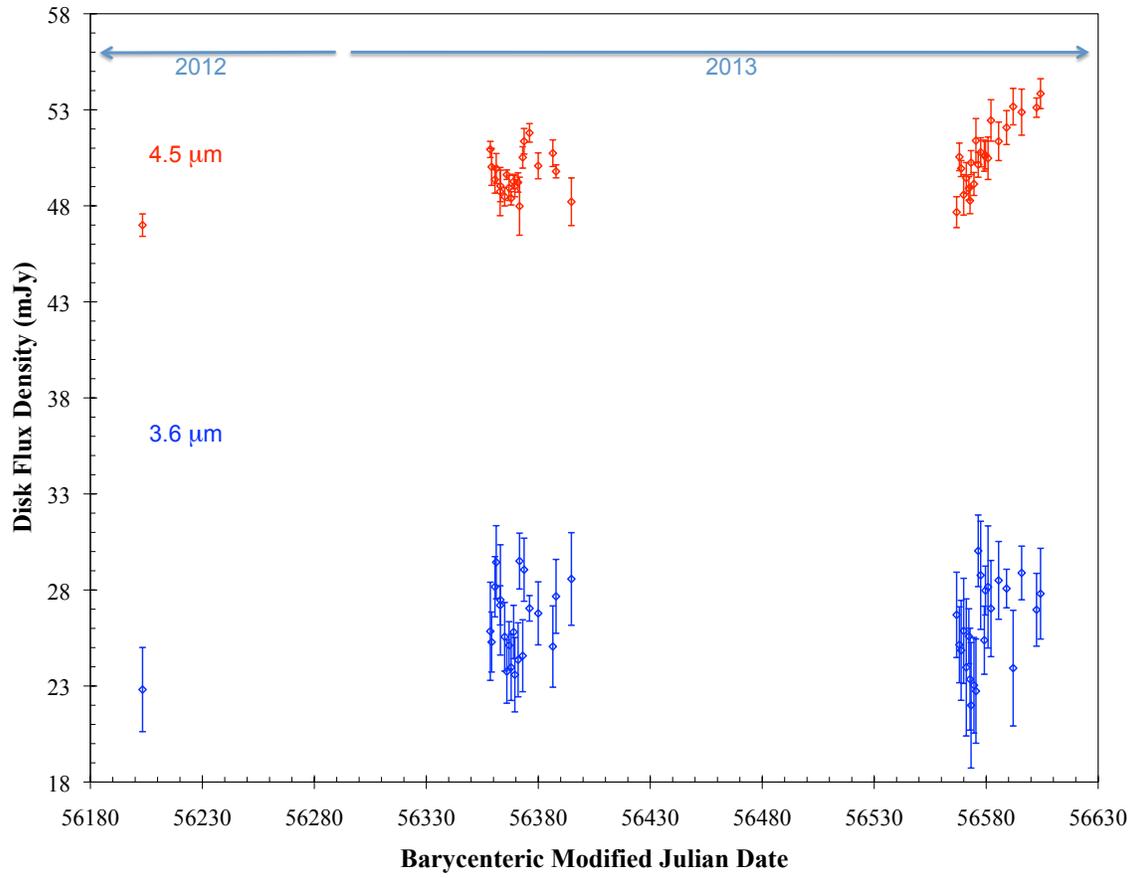}
\caption{Infrared light curve of the BD+20 307 disk in 2012 and 2013, showing no decay and significant growth trend at 4.5 $\micron$.\label{BD20307}}
\end{figure}

The infrared light curve of the disk of BD+20 307, as shown in Figure~\ref{BD20307}, did not show a decay from 2012 to 2013. Instead, the disk flux saw an upward trend at 4.5 $\micron$, while no significant tendency was seen at 3.6 $\micron$. With a linear approximation, we find that the average increase rate over the period was $0.4\pm1.1$ mJy per year at 3.6 $\micron$, and $2.5\pm0.7$ mJy per year at 4.5 $\micron$. The latter represents a few percent of the disk flux every year.

Although BD+20 307 has only two short visibility windows in a year, over the first 15 days of each visibility window in 2013, our infrared monitoring was conducted with a high average cadence of 1 AOR per day. Given a binary orbital period of 3.42 days, this sampling pattern was meant to cover $>$4 orbital cycles in an attempt to identify any associated effects in the disk. A SigSpec analysis of the growth line-detrended light curve reveals no traces of any periodicity. The strongest signals are at different periods at 3.6 and 4.5 $\micron$ with $sig$-values of only 3.00 and 2.37, respectively.

With the high cadence data, in this case we also tried the Plavchan periodogram \citep{pla08} as introduced in \S2.3. The algorithm reveals periods with significant power and apparently low false-alarm probability around 3.54 and 17.12 days at 3.6 $\micron$, and around 13.69 days at 4.5 $\micron$. The peak-to-peak amplitude is roughly 4 mJy, or 15\% of the average disk flux at 3.6 $\micron$, and 1.5 mJy or 3\% of the average disk flux at 4.5 $\micron$, not much greater than the nominal photometric errors at each waveband. The 3.54-day period is consistent with the binary orbital period. Since the Plavchan algorithm is sensitive to integer multiples of a period, the signals around 13.69 and 17.12 days are consistent with 4 and 5 times the binary orbital cycle of 3.42 days. More data with better time coverage are needed to confirm the periodicity of the disk output.

The color of the disk is $[3.6] - [4.5] = 1.19$, i.e., a blackbody temperature of 630 K. The $\chi^2$ test suggests that the color of neither the entire system nor the disk alone varied significantly with respect to the observational errors ($\chi_{red}^2 = 1.1$ in both cases).


\section{Comparisons and Implications}

\subsection{Incidence of terrestrial planets around solar-like stars}

Current wisdom is that terrestrial planets are present around a considerable fraction of solar-like stars \citep[e.g.,][]{fre13}. If their evolution is similar to that of the solar system, for each system we would expect to see the consequences of many impacts during the course of their formation \citep{ken05}. To order of magnitude, $\sim$10 giant impacts (between planetary embryos) are needed for the formation of an Earth-like planet \citep{ste12}. So in a system of multiple terrestrial planets, we would expect to see 10-100 giant impacts in the era of terrestrial planet formation, roughly from 30 to 100+ Myr. At the same time, there should be orders of magnitude more smaller scale impacts between planetesimals, and between planetesimals and embryos \citep{ste12}.

However, observationally it appears that extraterrestrial impacts are not that common. An unbiased {\it Spitzer} survey of young open clusters in the age range of 30 to 130 Myr found only 3 stars with extreme debris disks out of 250 solar-like stars, indicating an incidence of $\sim$1\% \citep{bal09}. A census of nearby solar-like stars in the same age range essentially yielded the same result \citep{mel10}. With the ordinary assumption of impact fragmentation of large bodies and dust production in subsequent collisional cascades, the low occurrence rate of extreme debris disks would imply a fraction of $<$10\% for rocky planet host stars \citep{jac12}, contradicting the Kepler result that shows $>$17\% of solar-like stars have at least one Earth-like planet on orbits smaller than 85 days in period \citep{fre13}.

Now, the time domain observations may suggest a solution to this discrepancy. After the impact observed in 2013 around ID8, the disk decayed with a timescale of $\sim$370 days \citep{men14}. Similarly, the most dramatic debris disk evolution observed so far was around the 10 Myr K2 dwarf TYC 8241-2652-1, in which case the disk faded by $\sim$30 times over a period less than 2 years \citep{mel12}. In this work, we see that P1121 (and possibly HD 23514) also have decay timescales on the same order of magnitude. Therefore, one to a few years seem to be a basic timescale over which extreme debris disks would be depleted in the absence of significant dust replenishment. If the aftermath of large impacts fades so quickly, such impacts have to be more common than previously thought in order to make up the $\sim$1\% incidence of extreme debris disks around solar-like stars. More sophisticated models to reconcile the occurrence rates of extreme debris disks and terrestrial planets would require better understanding of the scale of the impacts \citep[See the Supplementary Materials in][]{men14}. But the potentially higher rate of impacts is qualitatively consistent with the high incidence of terrestrial planets in mature extrasolar systems.

\subsection{Are disk composition and evolution correlated?}

One of the most intriguing questions is whether the composition of a debris disk is correlated with its time domain characteristics. In the case of extreme debris disks, the mineralogical composition reflects the dust production mechanism and differential evolution. SiO gas is expected to be short-lived in the aftermath of large hypervelocity impacts \citep[][C. Lisse et al. 2015, in preparation]{lis09,joh12b}. Solid dust grains rich in glassy material will condense from the gas \citep{joh14} and can be retained on bound orbits unless they are small enough to be blown out by stellar radiation pressure. Amorphous and crystalline silicates can have a number of origins but are probably not direct products of impacts.

A summary of the time domain characteristics of all our targets is given in Table~\ref{characteristics}. Among these targets, the disks of HD 15407A, HD 23514, and HD 145263 are silica-rich according to their mid-infrared spectra (C. Lisse et al. 2015, in preparation), while ID8, P1121, and BD+20 307 are ordinary silicate-dominant systems and are practically silica-free \cite[][N. Gorlova et al. 2015, in preparation]{wei11,olo12}. BD+20 307 is not considered in this comparison because it is too old. By $\gtrsim$1 Gyr, the star probably have had a mature planetary system. The emergence of a dusty disk at this age may be caused by some different process, and the disk is likely in a different dynamical environment than others in the era of terrestrial planet formation. However, a critical limitation with this comparison is that we do not have mid-infrared spectroscopic observations in 2012 and 2013. The available mineralogical characteristics are based on the IRS spectra obtained by the cryogenic {\it Spitzer} mission before May 2009. If there were major collisions during the gap that had led to significant mineralogical evolution in the disk, our comparison could be misleading.

All but one (HD 145263) of the young disks in our sample are variable at 3.6 and 4.5 $\micron$ in 2012-2013, but there appear to be noticeable differences between the silicate and silica disks. Both silicate disks in our sample, ID8 and P1121, have fluxes continuously decreasing by a large fraction of the total on timescales of 1-2 years. On the other hand, among the silica-rich disks, HD 15407A and HD 145263 did not show any decay in 2013, though HD 145263 did follow a weak downward trend over a $\sim$4 year interval by comparing multi-epoch IRS observations. The other silica-rich disk, HD 23514, showed slightly attenuated flux over a 1.1-year time baseline, but the decrease is too weak to be certainly ascribed to a monotonic decay.

In addition to our targets in this work, another example of a silica-rich disk is HD 172555, which is not an extreme debris disk ($f_d \sim 8\times10^{-4}$) but has a large amount of silica dust with mid-IR flux stable within 4\% over 27 years \citep{joh12b}. These examples appear to suggest that silica-rich systems do not vary as quickly as silicate-dominant systems. To test this hypothesis, more observations over longer time baselines will be needed to better characterize both the short-term behavior and long-term trend of the silica-rich disks. In particular, we may need to focus on the spectroscopic variations, or the wavelengths of silica/silicate emission bands, to get more sensitive monitoring of collision events.

Finally, we notice that the two types of disks may have different color-magnitude relations. The data of ID8 and P1121 are consistent with a constant stellar flux plus a disk with either varying temperature or varying dust emitting area, or their combination. By contrast, we did not see any significant color variations in the silica-rich disks, let alone any color-magnitude relation. As the temporal characteristics of the disks tend to reflect their dynamical evolution, the correlation between disk composition and time domain characteristics suggests that silicate-dominant and silica-rich disks may have different origins or be in different stages in their evolution sequence.


\subsection{What causes the rapid decay after an impact?}

Though the decay of one to a few years seems common in extreme debris disks, such rapid evolution is unexpected from the theoretical perspective. An explanation of this timescale is the condensation of impact-induced vapor and the consequent mutual collisions between the condensates \citep{men14}. The physics of impact cratering suggests that the impact-induced vapor should condense over several hours \citep{joh12a} into glassy silicate spherules with diverse forms \citep{war08,joh14}. The short condensation timescale should limit any gas damping effect to be dynamically negligible on the interplanetary scale. The typical condensate size sensitively depends on the conditions of the impact, especially the impact velocity, but ranges from about 10 $\micron$ to 1 mm for planetesimal- to terrestrial planet-sized impactors \citep{joh12a}.

To compute the decay timescale of the hypothesized condensate cloud, we attribute the $\micron$-sized and smaller particles seen in a disk SED to the daughter products of collisions between the original condensates from the impact vapor, and then convert the observed fractional luminosity to its equivalent with the original condensate sizes while preserving the disk mass. This ignores the mass loss of the disk, but should provide the right order of magnitude unless the relied SED is observed too many years after an impact. Considering the range and distribution of possible condensate sizes in \citet{joh12a}, we find that the time needed to destroy those condensates is in the range of 100 days to 10 years \citep{wya02,zuc12} around solar-like stars. The distinction between such collisions and canonical collisional cascades lies on the sizes of the feeding objects. The collisional cascades in regular debris disks are feeded by a population of planetesimals on the order of 1 - 100 km \citep{wya02,qui07}; in the decaying extreme debris disks, the upper size limit of the feeding objects may be restricted by the condensation physics to no larger than the mm scale. In a dense environment with high fractional luminosity, such small feeding grains can be quickly depleted, leading to the yearly decay of disk excess.

Alternatively, the huge amounts of small grains in the extreme debris disks may initiate a rapid decay of the disk, known as a collisional avalanche, which is a chain reaction triggered by the release of a large amount of fine dust particles in shattering events, and sustained for a period of time by the consequent breakup of ambient dust grains upon collisions with the fine dust particles in the process of being blown out \citep{art97,gri07}. This may be a viable explanation for the disappearing disk around TYC 8241-2652-1 \citep{mel12}. An effective collisional avalanche requires a dense debris disk in which an outflowing dust particle has a considerable probability of hitting another grain to produce more outflowing particles on its way out. Extreme debris disks, by their definition of fractional luminosity, all satisfy this criterion of density. Particularly, some conditions of extreme debris disks, including the significant populations of sub-$\micron$-sized particles, high indices of power law size distributions, small distances to the star, all tend to increase the dust area amplification factor in support of a prominent avalanche \citep{gri07}. Counting from the time of the initial release of small grains, we expect a delay of a few orbital periods before the diminished disk emission becomes evident \citep{gri07,mel12}. In the case of ID8, given an orbital period of $\sim$71 days \citep{men14}, the delay should be about $\sim$200 days. The orbital periods of other extreme debris disks are not well determined. Case-specific modeling will be necessary to confirm to which degree the collisional avalanche model matches the observed delay of some of the extreme debris disks.

\section{Conclusions}

Five extreme debris disks around solar-like stars were observed with {\it Spitzer}/IRAC at 3.6 and 4.5 $\micron$ in 2012 to 2013 (into 2014 for one source, P1121). All but one of the systems (BD+20 307) are in the age range from 10 to 120 Myr, roughly the era of violent collisions for terrestrial planet formation. All but one source (HD 145263) were also monitored in the optical at the same time from the ground. Together with the ID8 disk studied in previous work \citep{men14}, we found

1. The variability of extreme debris disks is common and intrinsic, not driven by stellar variations.

2. Without effective replenishment of fine dust, extreme debris disks may fade on timescales on the order of one to a few years, much shorter than previously expected from collisional cascades. The timescales are consistent with the result of intensive collisions and/or a collisional avalanche after the vaporization of rocky materials caused by recent large impacts and the consequent condensation.

3. Disk composition and temporal evolution appear to be correlated. Limited to very small sample sizes in both categories, silicate-dominant disks have significant trends over timescales of order a year, with significant color variations. In comparison, silica-rich disks appear to have more random variations with weak or no trend on yearly timescales and no significant color variations. The correlation may suggests different origins or evolution stages of the disks.

\section*{Acknowledgements}
We acknowledge with thanks the variable star observations from the AAVSO International Database contributed by observers worldwide and used in this research. The authors would like to thank Alycia Weinberger for providing the published IRS spectrum of BD+20 307, and Rui Wang for the help in illustration preparation. This work is based on observations made with the Spitzer Space Telescope, which is operated by the Jet Propulsion Laboratory, California Institute of Technology under a contract with NASA. This work was partially supported by contracts 1255094 and 1256424 from Caltech/JPL to the University of Arizona, and by NASA Grant NNX10AD38G.

{\it Facilities:} \facility{AAVSO, Spitzer (IRAC)}

\appendix

\section{Optical Monitoring}

\subsection{Observations and Data Reduction}

Optical monitoring of the targets was obtained to identify any potential stellar influence on debris disk variability, critical information to understand the driving forces of the variations. However, because of {\it Spitzer}'s pointing restrictions and its relative position with regard to the Earth, a typical target will be too close to the Sun and unobservable from the ground at least in one of the two {\it Spitzer} visibility windows in a year.

Ground-based optical monitoring of \objectname{P1121} was made with the 0.41-m PROMPT 5 robotic telescope at Cerro Tololo Inter-American Observatory in Chile, with a typical cadence of 2 - 4 observations throughout the night every night if weather permitted. The CCD had a scale of 0.60\arcsec\ pixel$^{-1}$ and a field of view of 10\arcmin. The pointing repeatability of the telescope was not perfect, virtually equivalent to random dithering. Science images were prepared by an automatic pipeline with bias, dark, and flat field corrections applied and WCS aligned. Aperture photometry was made on the science images with a radius of 5 pixels and sky annulus between 15 and 30 pixels. The optical monitoring of \objectname{HD 15407A}, \objectname{HD 23514}, and \objectname{BD+20 307} were supported by the American Association of Variable Star Observers (AAVSO), and conducted by different observers of the AAVSO. \objectname{HD 145263} was not monitored in the optical in 2013.

Due to our pursuit of intensive time coverage, not all ground-based observations were made under photometric weather. Therefore, differential photometry with respect to selected comparison stars was used as the general strategy for all optical observations. For each target, we also checked the photometry against at least 1 additional comparison star to avoid apparent variability caused by changes in the first comparison star. A summary of the observations is listed in Table~\ref{optical}. The optical observations show that all monitored stars are stable within the measurement uncertainties.

\subsection{P1121}

\begin{figure}
\epsscale{1}
\plotone{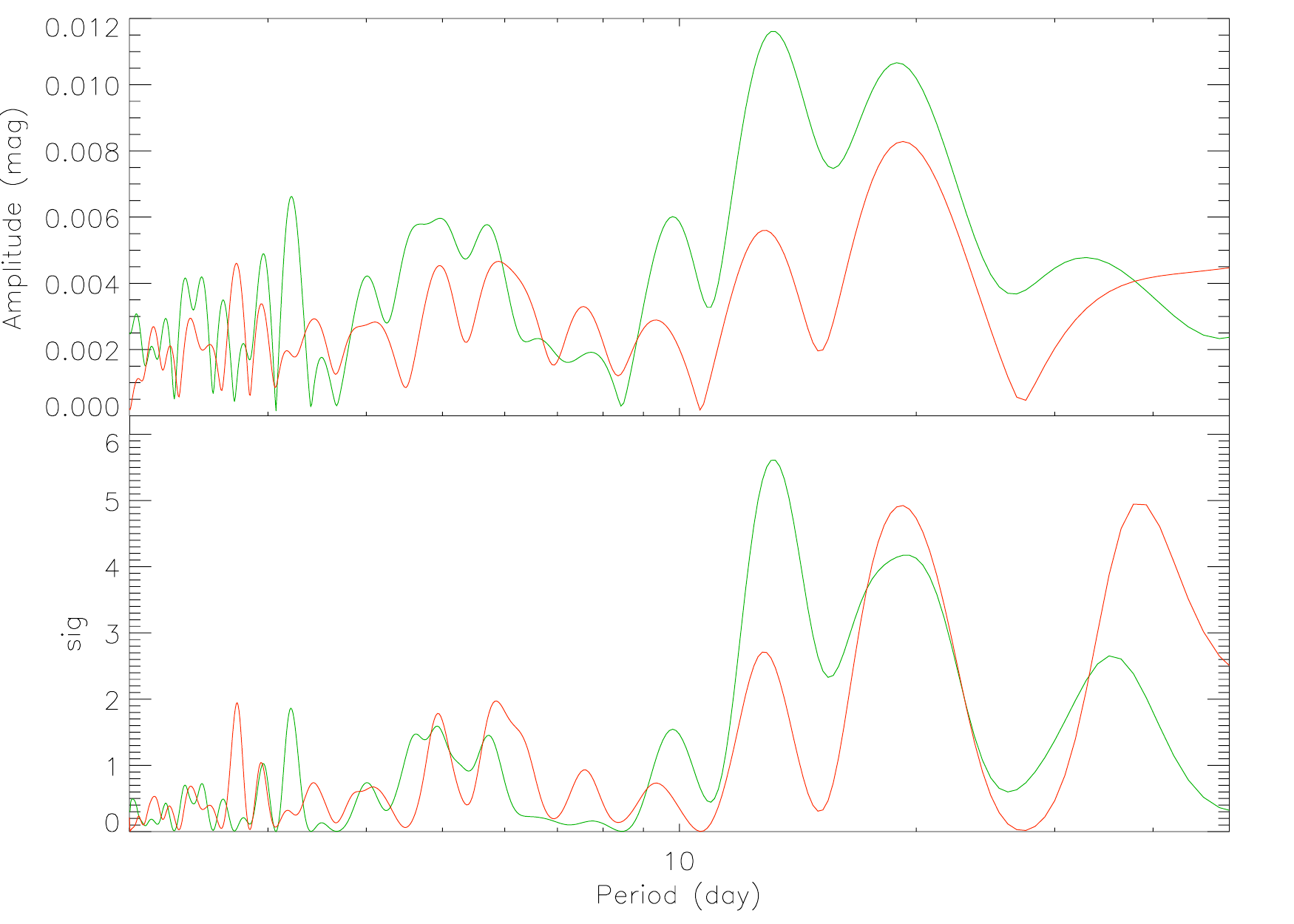}
\caption{Optical periodogram of P1121. The blue and red lines represent the data in $V$ and $I_C$ bands, respectively.\label{P1121_optical_fig}}
\end{figure}

The optical light curve of P1121 is flat, with a total RMS of 0.013 magnitude in $V$ and 0.009 magnitude in $I_C$. The data are given in Table~\ref{P1121_optical_tab}. As shown in Figure~\ref{P1121_optical_fig}, Fourier analysis reveals similar periodograms with apparent features near 13, 19, and 39 days in both wavebands, with significance right around our threshold. These detections are independent of the reference stars used to obtain the differential photometry. An evaluation with harmonics suggests that they possibly belong to an overtone system, of which the fundamental period, combining both $V$ and $I_C$ band data, is $39.0 \pm 10.5$ day. The large error is the result of the low amplitude compared to the photometric uncertainties, and from the short time baseline ($\sim$51 days) compared to the period \citep{kal08}. The signal may need to be confirmed by future observations.

Assuming the optical periods are real and are from stellar activity, they are too small to influence the disk changes seen in the infrared. The presence of harmonics in the optical is expected when the stellar surface has a complicated pattern of starspot distribution \citep{hul11}. But given the stellar radius of 1.14 R$_{\sun}$ of P1121, the optical period, if rotational, would be extraordinarily long for a solar-type star younger than 100 Myr.

Another possible explanation for the optical variations is debris disk veiling in front of the star. Though harmonics down to the second overtone are seen in the debris disk around ID8, they are found in direct observations because of a special geometric effect \citep{men14}. When it comes to transit timing, the debris disk may only impose the fundamental (orbital) period with no harmonics, unless the harmonics reflect a corresponding azimuthal density distribution of the disk. However, as the 19- and 39-day components are not far from the first and second infrared $ACF$ peaks between 15-20 days and around 35 days, this possibility cannot be ruled out with our current data set.

\subsection{HD 15407A}

Though the star was unobservable from the Earth during its first {\it Spitzer} visibility window, the second visibility window was completely covered by intensive optical observations, over a time baseline of $\sim$92 days. The data are given in Table~\ref{HD15407_optical_tab}.

\begin{figure}
\epsscale{1}
\plotone{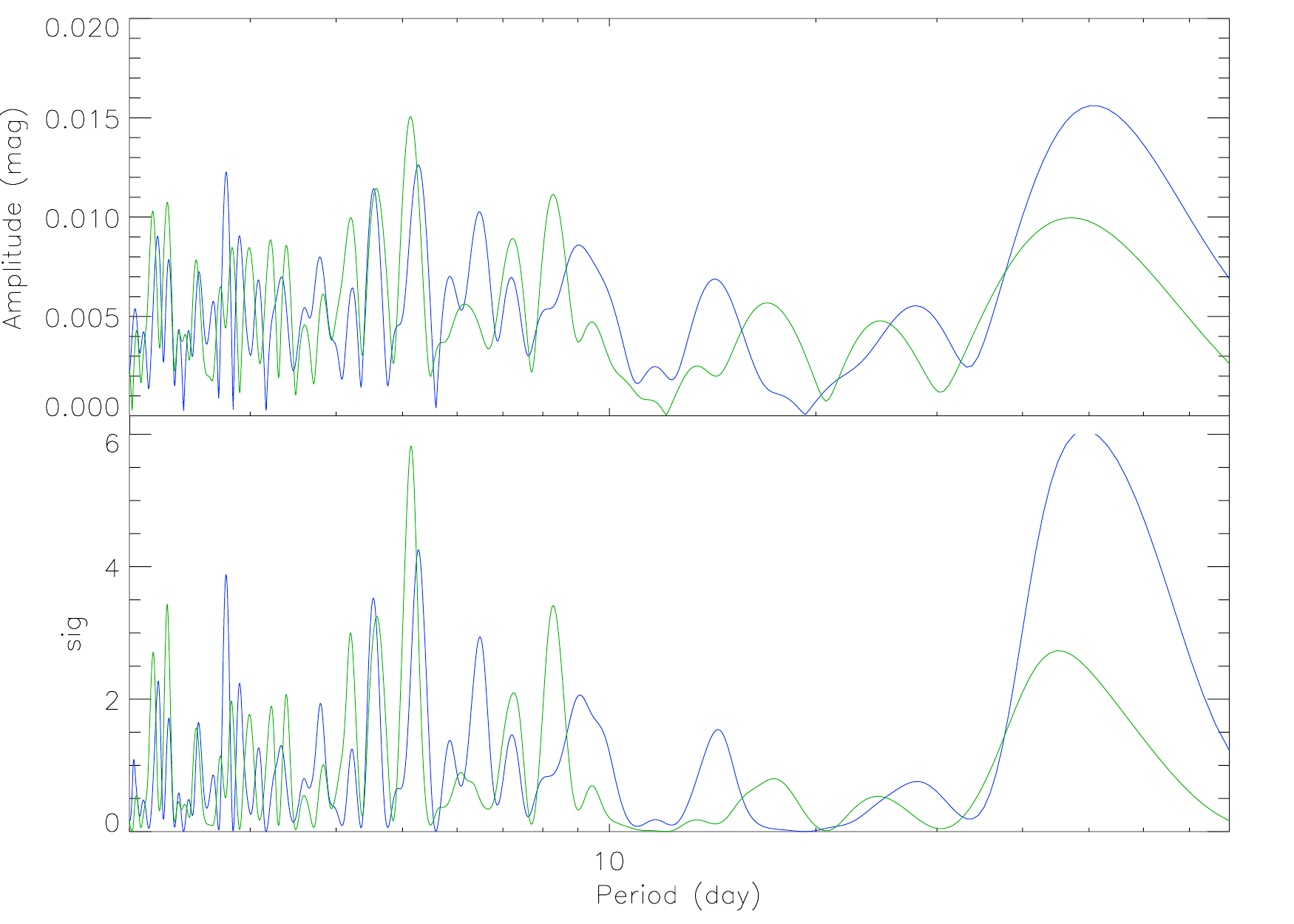}
\caption{Optical periodogram of HD 15407A. The blue and green lines represent the data in $B$ and $V$ bands, respectively.\label{HD15407A_optical}}
\end{figure}

The stellar light curve was stable throughout the observations. In the $B$ band periodogram, a peak near 50 days appears as the strongest signal. However, being longer than half the total time baseline, where red noise is readily masqueraded as broad peaks and the data become less constraining in folded phase curves, this period may be false and need additional confirmation. The next strongest signal is $\sim$5 days, which is significant ($sig = 5.05$) after removing the previous period. The same signal is the strongest in $V$ band at $sig = 5.83$. Combining the data from both wavebands, the period is determined at $5.20 \pm 0.09$ days, and is likely the rotational signature of the star (Figure~\ref{HD15407A_optical}). Given the stellar radius of 1.57 R$_{\sun}$, the equatorial rotational velocity of HD 15407A is $v = 15.3 \pm 0.3$ km s$^{-1}$. A spectroscopic measurement of the star with $\sim$6 km s$^{-1}$ resolution yields $v \sin i = 20$ km s$^{-1}$ \citep{mel10}. The two values, along with the fact that no traces of debris disk veiling are seen, mean that the star should be at an inclination close to 90\arcdeg and the disk nearly, but not exactly, edge-on.

\subsection{HD 23514AB}

The optical monitoring of HD 23514 in 2013 showed a flat light curve, with RMS of 0.027 magnitude in $B$, and 0.021 magnitude in $V$ band. The data are available in Table~\ref{HD23514_optical_tab}. Given that the observations were conducted by various observers with different telescopes, these values are consistent with the expected photometric errors. The brown dwarf companion should be fainter than the primary star by 13.6 and 12.9 magnitude in $B$ and $V$, respectively, and should have no influence on the monitoring. Fourier analysis fails to reveal any significant period in either optical band.

\subsection{BD+20 307AB}

The optical observations of BD+20 307 in 2013 revealed a flat light curve, with RMS of 0.018 magnitude in $V$. (The $B$ band observations are not considered because of the poor time coverage, Table~\ref{BD20307_optical_tab}.) The orbital period of the binary should be observable in the optical \citep{zuc08}, but we did not see it in the periodogram because of the photometric errors.

\clearpage

\clearpage

\clearpage

\clearpage

\clearpage

\clearpage

\clearpage

\clearpage

\clearpage

\clearpage

\clearpage

\clearpage

\clearpage

\clearpage

\clearpage

\LongTables

\begin{deluxetable}{cccccccc}
\tabletypesize{\scriptsize}
\tablecaption{Summary of {\it Spitzer} Observations\label{spitzer}}
\tablehead{
\colhead{} & \colhead{} & \multicolumn{2}{c}{Stellar flux (mJy)} & \colhead{} & \colhead{} & \colhead{} & \colhead{} \\
\cline{3-4}
\colhead{Source} & \colhead{Sp Type} & \colhead{3.6 $\micron$} & \colhead{4.5 $\micron$} & \colhead{Age (Myr)} & \colhead{Distance (pc)} & \colhead{Coverage (BMJD)} & \colhead{Cadence\tablenotemark{1} (day)}
}\\
\startdata
P1121 & F9 V & 9.35 & 6.05 & 80 & 490 & 56077.8 & \\
 & & & & & & 56311.2 - 56329.1 & 5 \\
 & & & & & & 56441.4 - 56484.0 & 5 \\
 & & & & & & 56657.0 - 56664.7 & 5 \\
HD 15407A & F5 V & 1259 & 806 & 80 & $55 \pm 2$ & 56371.1 - 56414.4 & 5 \\
 & & & & & & 56587.1 - 56635.9 & 5 \\
HD 23514A & F5 V & 152.2 & 98.6 & 120 & 130 & 56224.2 & \\
 & & & & & & 56388.8 - 56422.2 & 5 \\
 & & & & & & 56593.8 - 56632.0 & 5 \\
HD 23514B\tablenotemark{2} & M8 V & 0.405 & 0.292 & 120 & 130 & & \\
(within aperture) & & 0.348 & 0.246 & & & & \\
HD 145263 & F0 V & 193.3 & 125.0 & 10 & $150^{+26}_{-20}$ & 56416.3 - 56450.9 & 5 \\
 & & & & & & 56576.2 - 56614.4 & 5 \\
BD+20 307AB & G0 V & 259.2 & 162.3 & $\gtrsim$1000 & $96^{+13}_{-10}$ & 56203.3 & \\
 & & & & & & 56358.6 - 56394.8 & 4 (1) \\
 & & & & & & 56566.8 - 56604.3 & 4 (1) \\
\enddata
\tablenotetext{1}{Mean sampling interval. Numbers in parentheses denote for that in the high cadence phases, if there is any in that visibility window.}
\tablenotetext{2}{Not an intended target. Listed only for reference of the photometry of HD 23514A.}
\end{deluxetable}

\clearpage

\begin{deluxetable*}{cccc}
\tabletypesize{\footnotesize}
\tablecaption{{\it Spitzer}/IRAC Photometry of the Systems with Extreme Debris Disks\label{spitzer_phot}}
\tablehead{
\colhead{BMJD\tablenotemark{a}} & Waveband & \colhead{$F$ (mJy)} & $\sigma_{F}$ (mJy)
}
\startdata
\multicolumn{4}{c}{P1121}\\
\hline
56077.83283 & 3.6 $\micron$ & 11.737 & 0.059 \\
56311.19693 & 3.6 $\micron$ & 11.376 & 0.062 \\
56315.44559 & 3.6 $\micron$ & 11.070 & 0.065 \\
56318.87263 & 3.6 $\micron$ & 11.078 & 0.051 \\
56323.93743 & 3.6 $\micron$ & 11.027 & 0.033 \\
56329.12567 & 3.6 $\micron$ & 11.235 & 0.075 \\
56441.38477 & 3.6 $\micron$ & 10.732 & 0.055 \\
\enddata
\tablenotetext{a}{The mean Barycentric Modified Julian Date of the mid-time of all frames.}
\end{deluxetable*}

\clearpage

\begin{deluxetable}{cccccccc}
\tabletypesize{\scriptsize}
\tablecaption{Discriminants for Debris Disk Variability\label{variability}}
\tablehead{
\colhead{} & \colhead{} & \multicolumn{2}{c}{flux $\chi_{red}^2$} & \colhead{} & \multicolumn{2}{c}{color $\chi_{red}^2$} & \\
\cline{3-4} \cline{6-7}
\colhead{Source} & \colhead{$S$\tablenotemark{a}} & \colhead{3.6 $\micron$} & \colhead{4.5 $\micron$} & \colhead{} & \colhead{System} & \colhead{Disk} & \colhead{Slope (mJy yr$^{-1}$)}
}\\
\startdata
P1121 & 9.12 & 48 & 320 & & 7.7 & 3.8 & \\
HD 15407A & 0.81 & 0.7 & 2.1 & & 0.2 & 0.2 & \\
HD 23514A & 4.28 & 22 & 115 & & 3.3 & 5.4 & \\
HD 145263 & -0.28 & 3.0 & 1.4 & & 2.5 & 3.1 & $-3.2\pm2.3$ (3.6 $\micron$)  \\
 & & & & & & & $0.1\pm0.7$ (4.5 $\micron$) \\
BD+20 307AB & 0.47 & 1.2 & 5.3 & & 1.1 & 1.1 & $0.4\pm1.1$ (3.6 $\micron$) \\
 & & & & & & & $2.5\pm0.7$ (4.5 $\micron$) \\
\enddata
\tablenotetext{a}{Stetson index.}
\end{deluxetable}

\clearpage

\begin{deluxetable}{ccccccccc}
\tabletypesize{\scriptsize}
\tablecaption{Summary of the Time Domain Characteristics of the Samples\label{characteristics}}
\tablehead{
\colhead{} & \multicolumn{2}{c}{Silicate-Dominant Disk} & \colhead{} & \multicolumn{3}{c}{Silica-Rich Disk} & \colhead{} & \colhead{Old Disk} \\
\cline{2-3} \cline{5-7} \cline{9-9}
\colhead{Source} & \colhead{ID8} & \colhead{P1121} & \colhead{} & \colhead{HD 15047A} & \colhead{HD 23514A} & \colhead{HD 145263\tablenotemark{a}} & \colhead{} & \colhead{BD+20 307}
}\\
\startdata
Stellar Variations & N & N & & N & N & ? & & N \\
Disk Variations & Y & Y & & Y & Y & (Y) & & Y \\
Trend & Decay & Decay & & N & Long-term Decay(?) & (Long-term Decay) & & Growth \\
Periodicity & Y & Y(?) & & N & N & N & & Y(?) \\
Color Variations & Y & Y & & N & N & N & & N \\
Color Model\tablenotemark{b} & $\alpha$+$\beta$ & $\alpha$ & & $\gamma$ & $\gamma$ & $\gamma$ & & $\gamma$\\
\enddata
\tablenotetext{a}{Regular entries are based on the 3.6 and 4.5 $\micron$ IRAC observations. Entries in parentheses are based on the multi-epoch IRS spectroscopy.}
\tablenotetext{b}{Code for color variation models: $\alpha$ = temperature variations, $\beta$ = dust emitting area variations, $\gamma$ = no significant variations.}
\end{deluxetable}

\clearpage

\begin{table}
\begin{center}
\caption{Summary of Ground-based Optical Observations\label{optical}}
\scriptsize
\begin{tabular}{cccccc}
\tableline\tableline
 & & & & \multicolumn{2}{c}{Photometry Reference Stars} \\
\cline{5-6}
Source & Telescope & Coverage (BMJD) & Wavebands & Comparison & Check \\
\tableline
P1121 & CTIO 0.41-m PROMPT 5 & 56304.1 - 56317.1 & $V$, $I_C$ & P1225 & TYC 5409-601-1 \\
 & & 56334.1 - 56355.1 & & \\
HD 15407A & AAVSO Network & 56567.2 - 56659.2 & $B$, $V$ & HD 15358 & TYC 3691-1426-1 \\
 & & & & & HD 15407B \\
HD 23514AB & AAVSO Network & 56389.1 - 56399.1 & $V$ & TYC 1800-2018-1 & BD+22 552 \\
 & & 56525.4 - 56675.1 & $B$, $V$ & & \\
BD+20 307AB & AAVSO Network & 56559.4 - 56659.2  & $B$, $V$ & TYC 1212-374-1 & TYC 1212-226-1 \\
 & & & & & HR 577 \\
\tableline
\end{tabular}
\end{center}
\end{table}

\clearpage

\begin{deluxetable*}{ccccccc}
\tabletypesize{\footnotesize}
\tablecaption{Optical Monitoring of P1121\label{P1121_optical_tab}}
\tablehead{
\colhead{BMJD\tablenotemark{a}} & \colhead{Filter} & \colhead{Airmass} & \colhead{$m_1$\tablenotemark{b}} & \colhead{$\sigma_{m_1}$} & \colhead{$m_2$\tablenotemark{b}} & \colhead{$\sigma_{m_2}$}
}\\
\startdata
56304.09795 & $V$ & 1.295 & 0.568 & 0.002 & 0.553 & 0.002 \\
56304.09837 & $V$ & 1.293 & 0.574 & 0.002 & 0.559 & 0.002 \\
56305.09545 & $V$ & 1.294 & 0.579 & 0.002 & 0.565 & 0.002 \\
56305.09588 & $V$ & 1.292 & 0.606 & 0.002 & 0.574 & 0.002 \\
56306.12525 & $V$ & 1.153 & 0.575 & 0.002 & 0.550 & 0.002 \\
56306.12570 & $V$ & 1.151 & 0.574 & 0.002 & 0.552 & 0.002 \\
56307.08982 & $V$ & 1.297 & 0.582 & 0.002 & 0.558 & 0.002 \\
56307.09027 & $V$ & 1.294 & 0.581 & 0.002 & 0.552 & 0.002 \\
\enddata
\tablenotetext{a}{The Barycentric Modified Julian Date corresponding to the mid-time of an exposure.}
\tablenotetext{b}{Relative magnitudes with respect to the comparison star and check star listed in Table~\ref{optical}.}
\end{deluxetable*}

\clearpage

\begin{deluxetable*}{ccccccc}
\tabletypesize{\footnotesize}
\tablecaption{Optical Monitoring of HD 15407A\label{HD15407_optical_tab}}
\tablehead{
MJD\tablenotemark{a} & Filter & Airmass & $m$ & $\sigma_{m}$ & comp 1\tablenotemark{b} & comp 2\tablenotemark{b}
}\\
\startdata
56571.35949 & $B$ & 1.083 & 7.382 & 0.002 & HD 15358 & TYC 3691-1426-1 \\
56571.35972 & $B$ & 1.083 & 7.394 & 0.002 & HD 15358 & TYC 3691-1426-1 \\
56571.35991 & $B$ & 1.083 & 7.390 & 0.001 & HD 15358 & TYC 3691-1426-1 \\
56573.42546 & $B$ & 1.147 & 7.359 & 0.002 & HD 15358 & TYC 3691-1426-1 \\
56573.42563 & $B$ & 1.147 & 7.322 & 0.002 & HD 15358 & TYC 3691-1426-1 \\
56573.42579 & $B$ & 1.148 & 7.346 & 0.002 & HD 15358 & TYC 3691-1426-1 \\
56574.42106 & $B$ & 1.144 & 7.409 & 0.001 & HD 15358 & TYC 3691-1426-1 \\
56574.42123 & $B$ & 1.144 & 7.386 & 0.001 & HD 15358 & TYC 3691-1426-1 \\
\enddata
\tablenotetext{a}{The Modified Julian Date given in the AAVSO database.}
\tablenotetext{b}{The comparison star and check star. An entry is left blank if a check star is not used or if a star used is not in Table~\ref{optical}. See the AAVSO database for more details.}
\end{deluxetable*}

\clearpage

\begin{deluxetable*}{ccccccc}
\tabletypesize{\footnotesize}
\tablecaption{Optical Monitoring of HD 23514\label{HD23514_optical_tab}}
\tablehead{
MJD\tablenotemark{a} & Filter & Airmass & $m$ & $\sigma_{m}$ & comp 1\tablenotemark{b} & comp 2\tablenotemark{b}
}\\
\startdata
56558.32754 & $B$ & 1.130 & 9.921 & 0.004 & BD+22 552 & TYC 1800-2018-1 \\
56560.32155 & $B$ & 1.130 & 9.951 & 0.004 & BD+22 552 & TYC 1800-2018-1 \\
56561.31868 & $B$ & 1.130 & 10.206 & 0.004 & BD+22 552 & TYC 1800-2018-1 \\
56562.32681 & $B$ & 1.100 & 9.927 & 0.005 & BD+22 552 & TYC 1800-2018-1 \\
56563.31400 & $B$ & 1.130 & 9.950 & 0.004 & BD+22 552 & TYC 1800-2018-1 \\
56565.29818 & $B$ & 1.332 & 9.860 & 0.004 & TYC 1800-2018-1 & BD+22 552 \\
56565.29843 & $B$ & 1.330 & 9.871 & 0.004 & TYC 1800-2018-1 & BD+22 552 \\
56565.29866 & $B$ & 1.329 & 9.840 & 0.004 & TYC 1800-2018-1 & BD+22 552 \\
\enddata
\tablenotetext{a}{The Modified Julian Date given in the AAVSO database.}
\tablenotetext{b}{The comparison star and check star. An entry is left blank if a check star is not used or if a star used is not in Table~\ref{optical}. See the AAVSO database for more details.}
\end{deluxetable*}

\clearpage

\begin{deluxetable*}{ccccccc}
\tabletypesize{\footnotesize}
\tablecaption{Optical Monitoring of BD+20 307\label{BD20307_optical_tab}}
\tablehead{
MJD\tablenotemark{a} & Filter & Airmass & $m$ & $\sigma_{m}$ & comp 1\tablenotemark{b} & comp 2\tablenotemark{b}
}\\
\startdata
56593.06000 & $B$ & & 9.531 & 0.018 & & \\
56593.08300 & $B$ & & 9.529 & 0.014 & & \\
56593.10500 & $B$ & & 9.534 & 0.011 & & \\
56595.23200 & $B$ & & 9.530 & 0.003 & & \\
56607.03000 & $B$ & & 9.525 & 0.008 & & \\
56608.14790 & $B$ & & 9.531 & 0.008 & & \\
56609.03810 & $B$ & & 9.529 & 0.011 & & \\
56619.17650 & $B$ & & 9.526 & 0.008 & & \\
\enddata
\tablenotetext{a}{The Modified Julian Date given in the AAVSO database.}
\tablenotetext{b}{The comparison star and check star. An entry is left blank if a check star is not used or if a star used is not in Table~\ref{optical}. See the AAVSO database for more details.}
\end{deluxetable*}


\begin{thebibliography}{}
\bibitem[Agnor et al.(1999)]{agn99} Agnor, C.B., Canup, R.M. \& Levison, H.F. 1999, Icarus, 142, 219
\bibitem[Allard et al.(2011)]{all11} Allard, F., Homeier, D. \& Freytag, B. 2011, 16th Cambridge Workshop on Cool Stars, Stellar Systems, and the Sun (ASP Conf. Ser. 448), ed. C. Johns-Krull, M.K. Browning \& A.A. West (San Francisco, CA: ASP), 91
\bibitem[Artymowicz(1988)]{art88} Artymowicz, P. 1988, \apj, 335, L79
\bibitem[Artymowicz(1997)]{art97} Artymowicz, P. 1997, Annu. Rev. Earth Planet. Sci., 25, 175
\bibitem[Backman \& Paresce(1993)]{bac93} Backman, D.E. \& Paresce, F. 1993, in Protostars and Planets III, ed. E.H. Levy \& J.I. Lunine (Tucson, AZ: Univ. Arizona Press), 1253
\bibitem[Balog et al.(2009)]{bal09}Balog, Z., Kiss, L. L., Vink\'{o}, J, Rieke, G.H., Muzerolle, J., G\'{a}sp\'{a}r, A., Young, E.T., Gorlova, N, 2009, \apj, 698, 1989
\bibitem[Beichman et al.(2011)]{bei11} Beichman, C.A. et al. 2011, \apj, 743, 85
\bibitem[Canup(2004)]{can04}Canup, R. M. 2004, \araa, 42, 441
\bibitem[Canup(2012)]{can12}Canup, R. M. 2012, Science, 338, 1052
\bibitem[Castelli \& Kurucz(2004)]{cas04} Castelli, F. \& Kurucz, R.L. 2004, arXiv:astro-ph/1104.2185
\bibitem[Chambers(2013)]{cha13} Chambers, J.E. 2013, Icarus, 224, 43
\bibitem[Chen et al.(2014)]{che14} Chen, C.H., Mittal, T., Kuchner, M., Forrest, W.J., Lisse, C.M., Manoj, P., Sargent, B.A. \& Watson, D.M. 2014, \apjs, 211, 25
\bibitem[Cody et al.(2014)]{cod14} Cody, A.M. et al. 2014, \aj, 147, 82
\bibitem[\'{C}uk \& Stewart(2012)]{cuk12} \'{C}uk, M. \& Stewart, S.T. 2012, Science, 338, 1047
\bibitem[Dohnanyi (1969)]{doh69} Dohnanyi, J.S. 1969, \jgr, 74, 2531
\bibitem[Droege et al.(2006)]{dro06} Droege, T.F., Richmond, M.W., Sallman, M.P. \& Creager, R.P. 2006, \pasp, 118, 1666
\bibitem[Farihi et al.(2009)]{far09} Farihi, J., Jura, M. \& Zuckerman, B. 2009, \apj, 694, 805
\bibitem[Fazio et al.(2004)]{faz04} Fazio, G.G. et al. 2004, \apjs, 154, 10
\bibitem[Fekel et al.(2012)]{fek12} Fekel, F.C., Cordero, M.J., Galicher, R., Zuckerman, B., Melis, C. \& Weinberger, A.J. 2012, \apj, 749, 7
\bibitem[Flaherty et al.(2012)]{fla12} Flaherty, K.M., Muzerolle, J., Rieke, G., Gutermuth, R., Balog, Z., Herbst, W., Megeath, S.T. \& Kun, M. 2012, \apj, 748, 71
\bibitem[Flaherty et al.(2013)]{fla13} Flaherty, K.M., Muzerolle, J., Rieke, G., Gutermuth, R., Balog, Z., Herbst, W. \& Megeath, S.T. 2013, \aj, 145, 66
\bibitem[Fressin et al.(2013)]{fre13} Fressin, F. et al. 2013, \apj, 766, 81
\bibitem[Fujiwara et al.(2012a)]{fuj12a} Fujiwara, H., Onaka, T., Yamashita, T., Ishihara, D., Kataza, H., Fukagawa, M., Takeda, Y. \& Murakami, H. 2012a, \apj, 749, L29
\bibitem[Fujiwara et al.(2012b)]{fuj12b} Fujiwara, H., Onaka, T., Takita, S., Yamashita, T., Fukagawa, M., Ishihara, D., Kataza, H. \& Murakami, H. 2012b, \apj, 759, L18
\bibitem[G\'{a}sp\'{a}r et al.(2012)]{gas12} G\'{a}sp\'{a}r, A., Psaltis, D., Rieke, G.H. \& \"{O}zel, F. 2012, \apj, 754, 74
\bibitem[G\'{a}sp\'{a}r et al.(2013)]{gas13} G\'{a}sp\'{a}r, A., Rieke, G.H. \& Balog, Z. 2013, \apj, 768, 25
\bibitem[Gorlova et al.(2004)]{gor04} Gorlova, N. et al. 2004, \apjs, 154, 448
\bibitem[Gorlova et al.(2007)]{gor07} Gorlova, N., Balog, Z., Rieke, G.H., Muzerolle, J., Su, K.Y.L. 2007, \apj, 670, 516
\bibitem[Grigorieva et al.(2007)]{gri07} Grigorieva, A., Artymowicz, P. \& Th\'{e}bault, P. 2007, \aap, 461, 537
\bibitem[H{\o}g et al.(2000)]{hog00} H{\o}g, E. et al. 2000, \aap, 355, L27
\bibitem[Honda et al.(2003)]{hon03} Honda, M., Kataza, H., Okamoto, Y.K., Miyata, T., Yamashita, T., Sako, S., Takubo, S. \& Onaka, T. 2003, \apj, 585, L59
\bibitem[Honda et al.(2004)]{hon04} Honda, M. et al. 2004, \apj, 610, L49
\bibitem[Houck et al.(2004)]{hou04} Houch, J.R. et al. 2004, \apjs, 154, 18
\bibitem[Hulot et al.(2011)]{hul11} Hulot, J.C., Baudin, F., Samadi, R. \& Goupil, M.J. 2011, arXiv:1104.2185
\bibitem[Ida et al.(1997)]{ida97} Ida, S., Canup, R.M. \& Stewart, G.R. 1997, \nat, 389, 353
\bibitem[Jackson \& Wyatt(2012)]{jac12} Jackson, A.P. \& Wyatt, M.C. 2012, \mnras, 425, 657
\bibitem[Janson et al.(2013)]{jan13} Janson, M., Lafreni\`{e}re, D., Jayawardhana, R., Bonavita, M., Girard, J.H., Brandeker., A. \& Gizis, J.E. 2013, \apj, 773, 170
\bibitem[Johnson \& Melosh(2012)]{joh12a} Johnson, B.C. \& Melosh, H.J. 2012, \icarus, 217, 416
\bibitem[Johnson \& Melosh(2014)]{joh14} Johnson, B.C. \& Melosh, H.J. 2014, \icarus, 228, 347
\bibitem[Johnson et al.(2012)]{joh12b} Johnson, B.C. 2012, \apj, 761, 45
\bibitem[Kallinger et al.(2008)]{kal08} Kallinger, T., Reegen, P. \& Weiss, W.W. 2008, \aap, 481, 571
\bibitem[Kenyon \& Bromley(2005)]{ken05} Kenyon, S.J. \& Bromley, B.C. 2005, \aj, 130, 269
\bibitem[Kenyon \& Bromley(2006)]{ken06} Kenyon, S.J. \& Bromley, B.C. 2006, \aj, 131, 1837
\bibitem[Kokubo \& Ida(2006)]{kok07} Kokubo, E. \& Ida, S. 2007, \apj, 671, 2082
\bibitem[Lebouteiller et al.(2011)]{leb11} Lebouteiller, V., Barry, D.J., Spoon, H.W.W., Bernard-Salas, J., Sloan, G.C., Houck, J.R., \& Weedman, D., 2011, \apjs, 196, 8
\bibitem[Lisse et al.(2009)]{lis09} Lisse, C.M., Chen, C.H., Wyatt, M.C., Morlok, A., Song, I., Gryden, G. \& Sheehan, P. 2009, \apj, 701, 2019
\bibitem[Melis et al.(2010)]{mel10} Melis, C., Zuckerman, B., Rhee, J.H., Song, I. 2010, \apj, 717, L57
\bibitem[Melis et al.(2012)]{mel12} Melis, C., Zuckerman, B., Rhee, J.H., Song, I., Murphy, S.J., Bessell, M.S. 2012, \nat, 487, 74
\bibitem[Meng et al.(2012)]{men12} Meng, H.Y.A., Rieke, G.H., Su, K.Y.L., Ivanov, V.D., Vanzi, L. \& Rujopakarn, W. 2012, \apj, 751, L17
\bibitem[Meng et al.(2014)]{men14} Meng, H.Y.A. et al. 2014, Science, 345, 1032
\bibitem[Morales et al.(2011)]{mor11} Morales, F.Y. Rieke, G.H., Werner, M.W., Bryden, G., Stapelfeldt, K.R. \& Su, K.Y.L. 2008, \apj, 730, L29
\bibitem[Morishima et al.(2010)]{mor10} Morishima, R., Stadel, J. \& Moore, B. 2010, Icarus, 207, 517
\bibitem[Morlok et al.(2014)]{mor14} Morlok, A., Mason, A.B., Anand, M., Lisse, C.M., Bullock, E.S. \& Grady, M.M. 2014, Icarus, 239, 1
\bibitem[Ofek(2008)]{ofe08} Ofek, E.O. 2008, \pasp, 120, 1128
\bibitem[Olofsson et al.(2012)]{olo12} Olofsson, J., Juh\'{a}sz, A., Henning, T., Mutschke, H., Tamanai, A., Mo\'{o}r, A. \& \'{A}brah\'{a}m, P. 2012, \aap, 542, 90
\bibitem[Oudmaijer et al.(1992)]{oud92} Oudmaijer, R.D., van der Veen, W.E.C.J., Waters, L.B.F.M., Trams, N.R., Waelkens, C., Engelsman, E. 1992, \aaps, 96, 625
\bibitem[Pecaut et al.(2012)]{pec12} Pecaut, M.J., Mamajek, E.E. \& Bubar, E.J. 2012, \apj, 746, 154
\bibitem[Perryman et al.(1997)]{per97} Perryman, M.A.C. et al. 1997, \aap, 323, 49
\bibitem[Plavchan et al.(2008)]{pla08} Plavchan, P., Jura, M., Kirkpatrick, J.D., Cutri, R.M. \& Gallagher, S.C. 2008, \apjs, 175, 191
\bibitem[Prisinzano et al.(2003)]{pri03} Prisinzano, L., Micela, G., Sciortino, S. \& Favata, F. 2003, \aap, 404, 927
\bibitem[Quillen et al.(2007)]{qui07} Quillen, A.C., Morbidelli, A. \& Moore, A. 2003, \mnras, 380, 1642
\bibitem[Raymond et al.(2014)]{ray14} Raymond, S.N., Kokubo, E., Morbidelli, A., Morishima, R. \& Walsh, K.J. 2014, in Protostars and Planets VI, ed. H. Beuther, R. Klessen, C. Dullemond \& T. Henning (Tucson, AZ: Univ. Arizona Press), in press
\bibitem[Rebull et al.(2014)]{reb14} Rebull, L. et al. 2014, \aj, 148, 92
\bibitem[Reegen(2007)]{ree07} Reegen, P. 2007, \aap, 467, 1353
\bibitem[Rhee et al.(2008)]{rhe08} Rhee, J.H., Song, I., Zuckerman, B. 2008, \apj, 675, 777
\bibitem[Rieke et al.(2004)]{rie04} Rieke, G.H. et al., 2004, \apjs, 154, 25
\bibitem[Rieke et al.(2005)]{rie05} Rieke, G.H. et al., 2005, \apj, 620, 1010
\bibitem[Righter \& O'Brien(2011)]{rig11}Righter, K. \& O'Brien, D.P. 2011, Proc. of Nat. Acad. Sci., 108, 19165
\bibitem[Rodriguez et al.(2012)]{rod12} Rodriguez, D.R., Marois, C., Zuckerman, B., Macintosh, B. \& Melis, C. 2012, \apj, 748, 30
\bibitem[Rojo Arellano et al.(1997)]{roj97} Rojo Arellano, E., Pe\~{n}a, J. \& Gonz\'{a}lez, D. 1997, \aaps, 123, 25
\bibitem[Schneider et al.(2013)]{sch13} Schneider, A., Song, I., Melis, C., Zuckerman, B., Bessell, M., Hufford, T. \& Hinkley, S. 2013, \apj, 777, 78
\bibitem[Skrutskie et al.(2006)]{skr06} Skrutskie, M.F. et al. 2006, \aj, 131, 1163
\bibitem[Soderblom et al.(2014)]{sod14} Soderblom, D.R., Hillenbrand, L.A., Jeffries, R.D., Mamajek, E.E. \& Naylor, T. 2014, in Protostars and Planets VI, ed. H. Beuther, R. Klessen, C. Dullemond \& T. Henning (Tucson, AZ: Univ. Arizona Press), in press
\bibitem[Song et al.(2005)]{son05} Song, I., Zuckerman, B., Weinberger, A.J. \& Becklin, E.E. 2005, \nat, 436, 363
\bibitem[Spitzer \& Kleinman(1961)]{spi61} Spitzer, W.G. \& Kleinman, D.A. 1961, PhRv, 121, 1324
\bibitem[Stetson(1996)]{ste96} Stetson, P.B. 1996, \pasp, 108, 851
\bibitem[Stewart \& Leinhardt(2012)]{ste12} Stewart, S.T. \& Leinhardt, Z.M. 2012, \apj, 751, 32
\bibitem[Su et al.(2006)]{su06} Su, K.Y.L. et al. 2006, \apj, 653, 675
\bibitem[Takasawa et al.(2011)]{tak11} Takasawa, S. et al. 2011, \apj, 733, L39
\bibitem[Warren(2008)]{war08} Warren, P.H. 2008, \gca, 72, 3562
\bibitem[Weinberger(2008)]{wei08} Weinberger, A.J. 2008, \apj, 679, L41
\bibitem[Weinberger et al.(2011)]{wei11} Weinberger, A.J., Becklin, E.E. \& Zuckerman, B. 2011, \apj, 726, 72
\bibitem[Wyatt \& Dent(2002)]{wya02} Wyatt, M.C. \& Dent, W.R.F. 2002, \mnras, 334, 589
\bibitem[Wyatt et al.(2007)]{wya07} Wyatt, M.C., Smith, R., Greaves, J.S., Beichman, C.A., Bryden, G. \& Lisse, C.M. 2007, \apj, 658, 569
\bibitem[Wyatt(2008)]{wya08} Wyatt, M.C. 2008, \araa, 46, 339
\bibitem[Yamamoto et al.(2013)]{yam13} Yamamoto, K. et al. 2013, \pasj, 65, 90
\bibitem[Zuckerman et al.(2008)]{zuc08} Zuckerman, B., Fekel, F.C., Williamson, M.H., Henry, G.W. \& Muno, M.P. 2008, \apj, 688, 1345
\bibitem[Zuckerman \& Song(2012)]{zuc12} Zuckerman, B. \& Song, I. 2012, \apj, 758, 77
\end{thebibliography}
\end{document}